\documentclass[superscriptaddress,reprint,amsmath,amssymb,aps,prd,showpacs,nofootinbib]{revtex4-1}
\usepackage{units}
\usepackage{graphicx}
\usepackage{dcolumn}
\usepackage{subfigure}  
\usepackage{bm}
\usepackage{xspace}
\usepackage{soul}
\usepackage{cancel}
\usepackage{tabularx}
\usepackage{color}
\usepackage{multirow} 
\usepackage[demo,abs]{overpic}
\usepackage[mathlines]{lineno}

\begin{document}
\renewcommand{\arraystretch}{1.2}
\newcommand{\zeronu}{0$\nu\beta\beta$\xspace}
\newcommand{\twonu}{2$\nu\beta\beta$\xspace}
\newcommand{\nemo}{NEMO-3\xspace}
\newcommand{\mbb}{$\langle m_{\nu} \rangle$\xspace}
\newcommand{\changetext}[2]{{\color{blue}\st{#1} #2}}
\newcommand{\addtext}[1]{{\color{red}(\emph{Added text}) #1}}
\newcommand{\removetext}[1]{{\color{red}\st{#1}}}
\newcommand{\movetext}[1]{{\color{blue}#1}}
\def \nobreakseq {\nobreak \hskip 0pt \hbox}

\title{Measurement of the 2$\nu\beta\beta$ decay half-life of $^{150}$Nd and
 a search for 0$\nu\beta\beta$ decay processes with the full exposure from the NEMO-3 detector}

\author{R.~Arnold}
\affiliation{IPHC, ULP, CNRS/IN2P3\nobreakseq{,} F-67037 Strasbourg, France}
\author{C.~Augier} 
\affiliation{LAL, Universit\'{e} Paris-Sud\nobreakseq{,} CNRS/IN2P3\nobreakseq{,} Universit\'{e} Paris-Saclay\nobreakseq{,}
F-91405 Orsay\nobreakseq{,} France}
\author{J.D.~Baker}
\thanks{Deceased}
\affiliation{Idaho National Laboratory\nobreakseq{,} Idaho Falls, ID 83415, U.S.A.}
\author{A.S.~Barabash}
\affiliation{NRC ``Kurchatov Institute", ITEP, 117218 Moscow, Russia}
\author{A.~Basharina-Freshville} 
\affiliation{UCL, London WC1E 6BT\nobreakseq{,} United Kingdom}
\author{S.~Blondel} 
\affiliation{LAL, Universit\'{e} Paris-Sud\nobreakseq{,} CNRS/IN2P3\nobreakseq{,} Universit\'{e} Paris-Saclay\nobreakseq{,}
F-91405 Orsay\nobreakseq{,} France}
\author{S.~Blot}
\affiliation{University of Manchester\nobreakseq{,} Manchester M13 9PL\nobreakseq{,}~United Kingdom}
\author{M.~Bongrand} 
\affiliation{LAL, Universit\'{e} Paris-Sud\nobreakseq{,} CNRS/IN2P3\nobreakseq{,} Universit\'{e} Paris-Saclay\nobreakseq{,}
F-91405 Orsay\nobreakseq{,} France}
\author{V.~Brudanin} 
\affiliation{JINR, 141980 Dubna, Russia}
\affiliation{National Research Nuclear University MEPhI, 115409 Moscow, Russia}
\author{J.~Busto} 
\affiliation{CPPM, Universit\'e de Marseille\nobreakseq{,} CNRS/IN2P3\nobreakseq{,} F-13288 Marseille\nobreakseq{,} France}
\author{A.J.~Caffrey}
\affiliation{Idaho National Laboratory\nobreakseq{,} Idaho Falls, ID 83415, U.S.A.}
\author{S. Calvez}
\affiliation{LAL, Universit\'{e} Paris-Sud\nobreakseq{,} CNRS/IN2P3\nobreakseq{,} Universit\'{e} Paris-Saclay\nobreakseq{,}
F-91405 Orsay\nobreakseq{,} France}
\author{M.~Cascella} 
\affiliation{UCL, London WC1E 6BT\nobreakseq{,} United Kingdom}
\author{C.~Cerna} 
\affiliation{CENBG\nobreakseq{,} Universit\'e de Bordeaux\nobreakseq{,} CNRS/IN2P3\nobreakseq{,} F-33175 Gradignan\nobreakseq{,} France}
\author{J.P.~Cesar}
\affiliation{University of Texas at Austin\nobreakseq{,}
  Austin\nobreakseq{,} TX 78712\nobreakseq{,}~U.S.A.}
\author{A.~Chapon} 
\affiliation{LPC Caen\nobreakseq{,} ENSICAEN\nobreakseq{,} Universit\'e de
Caen\nobreakseq{,} CNRS/IN2P3\nobreakseq{,} F-14050 Caen\nobreakseq{,} France}
\author{E.~Chauveau} 
\affiliation{University of Manchester\nobreakseq{,} Manchester M13 9PL\nobreakseq{,}~United Kingdom}
\author{A.~Chopra} 
\affiliation{UCL, London WC1E 6BT\nobreakseq{,} United Kingdom}
\author{D.~Duchesneau} 
\affiliation{LAPP, Universit\'e de Savoie\nobreakseq{,}
CNRS/IN2P3\nobreakseq{,} F-74941 Annecy-le-Vieux\nobreakseq{,} France}
\author{D.~Durand} 
\affiliation{LPC Caen\nobreakseq{,} ENSICAEN\nobreakseq{,} Universit\'e de
Caen\nobreakseq{,} CNRS/IN2P3\nobreakseq{,} F-14050 Caen\nobreakseq{,} France}
\author{V.~Egorov}
\affiliation{JINR, 141980 Dubna, Russia}
\author{G.~Eurin} 
\affiliation{LAL, Universit\'{e} Paris-Sud\nobreakseq{,} CNRS/IN2P3\nobreakseq{,} Universit\'{e} Paris-Saclay\nobreakseq{,}
F-91405 Orsay\nobreakseq{,} France}
\affiliation{UCL, London WC1E 6BT\nobreakseq{,} United Kingdom}
\author{J.J.~Evans} 
\affiliation{University of Manchester\nobreakseq{,} Manchester M13 9PL\nobreakseq{,}~United Kingdom}
\author{L.~Fajt} 
\affiliation{Institute of Experimental and Applied Physics\nobreakseq{,} Czech Technical University in Prague\nobreakseq{,} CZ-12800 Prague\nobreakseq{,} Czech Republic}
\author{D.~Filosofov} 
\affiliation{JINR, 141980 Dubna, Russia}
\author{R.~Flack} 
\affiliation{UCL, London WC1E 6BT\nobreakseq{,} United Kingdom}
\author{X.~Garrido} 
\affiliation{LAL, Universit\'{e} Paris-Sud\nobreakseq{,} CNRS/IN2P3\nobreakseq{,} Universit\'{e} Paris-Saclay\nobreakseq{,}
F-91405 Orsay\nobreakseq{,} France}
\author{H.~G\'omez} 
\affiliation{LAL, Universit\'{e} Paris-Sud\nobreakseq{,} CNRS/IN2P3\nobreakseq{,} Universit\'{e} Paris-Saclay\nobreakseq{,}
F-91405 Orsay\nobreakseq{,} France}
\author{B.~Guillon} 
\affiliation{LPC Caen\nobreakseq{,} ENSICAEN\nobreakseq{,} Universit\'e de
Caen\nobreakseq{,} CNRS/IN2P3\nobreakseq{,} F-14050 Caen\nobreakseq{,} France}
\author{P.~Guzowski} 
\affiliation{University of Manchester\nobreakseq{,} Manchester M13 9PL\nobreakseq{,}~United Kingdom}
\author{R.~Hod\'{a}k} 
\affiliation{Institute of Experimental and Applied Physics\nobreakseq{,} Czech
Technical University in Prague\nobreakseq{,} CZ-12800
Prague\nobreakseq{,} Czech Republic}
\author{A.~Huber} 
\affiliation{CENBG\nobreakseq{,} Universit\'e de Bordeaux\nobreakseq{,} CNRS/IN2P3\nobreakseq{,} F-33175 Gradignan\nobreakseq{,} France}
\author{P.~Hubert} 
\affiliation{CENBG\nobreakseq{,} Universit\'e de Bordeaux\nobreakseq{,} CNRS/IN2P3\nobreakseq{,} F-33175 Gradignan\nobreakseq{,} France}
\author{C.~Hugon}
\affiliation{CENBG\nobreakseq{,} Universit\'e de Bordeaux\nobreakseq{,} CNRS/IN2P3\nobreakseq{,} F-33175 Gradignan\nobreakseq{,} France}
\author{S.~Jullian} 
\affiliation{LAL, Universit\'{e} Paris-Sud\nobreakseq{,} CNRS/IN2P3\nobreakseq{,} Universit\'{e} Paris-Saclay\nobreakseq{,}
F-91405 Orsay\nobreakseq{,} France}
\author{A.~Klimenko} 
\affiliation{JINR, 141980 Dubna, Russia}
\author{O.~Kochetov} 
\affiliation{JINR, 141980 Dubna, Russia}
\author{S.I.~Konovalov} 
\affiliation{NRC ``Kurchatov Institute", ITEP, 117218 Moscow, Russia}
\author{V.~Kovalenko}
\affiliation{JINR, 141980 Dubna, Russia}
\author{D.~Lalanne} 
\affiliation{LAL, Universit\'{e} Paris-Sud\nobreakseq{,} CNRS/IN2P3\nobreakseq{,} Universit\'{e} Paris-Saclay\nobreakseq{,}
F-91405 Orsay\nobreakseq{,} France}
\author{K.~Lang} 
\affiliation{University of Texas at Austin\nobreakseq{,}
  Austin\nobreakseq{,} TX 78712\nobreakseq{,}~U.S.A.}
\author{Y.~Lemi\`ere} 
\affiliation{LPC Caen\nobreakseq{,} ENSICAEN\nobreakseq{,} Universit\'e de
Caen\nobreakseq{,} CNRS/IN2P3\nobreakseq{,} F-14050 Caen\nobreakseq{,} France}
\author{T.~Le~Noblet} 
\affiliation{LAPP, Universit\'e de Savoie\nobreakseq{,} CNRS/IN2P3\nobreakseq{,} F-74941 Annecy-le-Vieux\nobreakseq{,} France}
\author{Z.~Liptak} 
\affiliation{University of Texas at Austin\nobreakseq{,}
  Austin\nobreakseq{,} TX 78712\nobreakseq{,}~U.S.A.}
\author{X.~R.~Liu} 
\affiliation{UCL, London WC1E 6BT\nobreakseq{,} United Kingdom}  
\author{P.~Loaiza} 
\affiliation{LAL, Universit\'{e} Paris-Sud\nobreakseq{,} CNRS/IN2P3\nobreakseq{,} Universit\'{e} Paris-Saclay\nobreakseq{,}
F-91405 Orsay\nobreakseq{,} France}
\author{G.~Lutter} 
\affiliation{CENBG\nobreakseq{,} Universit\'e de Bordeaux\nobreakseq{,} CNRS/IN2P3\nobreakseq{,} F-33175 Gradignan\nobreakseq{,} France}
\author{F.~Mamedov}
\affiliation{Institute of Experimental and Applied Physics\nobreakseq{,} Czech
Technical University in Prague\nobreakseq{,} CZ-12800
Prague\nobreakseq{,} Czech Republic}
\author{C.~Marquet} 
\affiliation{CENBG\nobreakseq{,} Universit\'e de Bordeaux\nobreakseq{,} CNRS/IN2P3\nobreakseq{,} F-33175 Gradignan\nobreakseq{,} France}
\author{F.~Mauger} 
\affiliation{LPC Caen\nobreakseq{,} ENSICAEN\nobreakseq{,} Universit\'e de
Caen\nobreakseq{,} CNRS/IN2P3\nobreakseq{,} F-14050 Caen\nobreakseq{,} France}
\author{B.~Morgan} 
\affiliation{University of Warwick\nobreakseq{,} Coventry CV4
7AL\nobreakseq{,} United Kingdom}
\author{J.~Mott} 
\affiliation{UCL, London WC1E 6BT\nobreakseq{,} United Kingdom}
\author{I.~Nemchenok} 
\affiliation{JINR, 141980 Dubna, Russia}
\author{M.~Nomachi} 
\affiliation{Osaka University\nobreakseq{,} 1-1 Machikaney arna
Toyonaka\nobreakseq{,} Osaka 560-0043\nobreakseq{,} Japan}
\author{F.~Nova} 
\affiliation{University of Texas at Austin\nobreakseq{,}
  Austin\nobreakseq{,} TX 78712\nobreakseq{,}~U.S.A.}
\author{F.~Nowacki} 
\affiliation{IPHC, ULP, CNRS/IN2P3\nobreakseq{,} F-67037 Strasbourg, France}
\author{H.~Ohsumi} 
\affiliation{Saga University\nobreakseq{,} Saga 840-8502\nobreakseq{,}
  Japan}
\author{R.B.~Pahlka}
\affiliation{University of Texas at Austin\nobreakseq{,}
  Austin\nobreakseq{,} TX 78712\nobreakseq{,}~U.S.A.}
\author{F.~Perrot} 
\affiliation{CENBG\nobreakseq{,} Universit\'e de Bordeaux\nobreakseq{,} CNRS/IN2P3\nobreakseq{,} F-33175 Gradignan\nobreakseq{,} France}
\author{F.~Piquemal} 
\affiliation{CENBG\nobreakseq{,} Universit\'e de Bordeaux\nobreakseq{,} CNRS/IN2P3\nobreakseq{,} F-33175 Gradignan\nobreakseq{,} France}
\affiliation{Laboratoire Souterrain de Modane\nobreakseq{,} F-73500
Modane\nobreakseq{,} France}
\author{P.~Povinec}
\affiliation{FMFI,~Comenius~Univ.\nobreakseq{,}~SK-842~48~Bratislava\nobreakseq{,}~Slovakia}
\author{P.~P\v{r}idal} 
\affiliation{Institute of Experimental and Applied Physics\nobreakseq{,} Czech Technical University in Prague\nobreakseq{,} CZ-12800 Prague\nobreakseq{,} Czech Republic}
\author{Y.A.~Ramachers} 
\affiliation{University of Warwick\nobreakseq{,} Coventry CV4
7AL\nobreakseq{,} United Kingdom}
\author{A.~Remoto}
\affiliation{LAPP, Universit\'e de Savoie\nobreakseq{,}
CNRS/IN2P3\nobreakseq{,} F-74941 Annecy-le-Vieux\nobreakseq{,} France}
\author{J.L.~Reyss} 
\affiliation{LSCE\nobreakseq{,} CNRS\nobreakseq{,} F-91190
  Gif-sur-Yvette\nobreakseq{,} France}
\author{B.~Richards} 
\affiliation{UCL, London WC1E 6BT\nobreakseq{,} United Kingdom}
\author{C.L.~Riddle} 
\affiliation{Idaho National Laboratory\nobreakseq{,} Idaho Falls, ID 83415, U.S.A.}
\author{E.~Rukhadze} 
\affiliation{Institute of Experimental and Applied Physics\nobreakseq{,} Czech
Technical University in Prague\nobreakseq{,} CZ-12800
Prague\nobreakseq{,} Czech Republic}
\author{R.~Saakyan} 
\affiliation{UCL, London WC1E 6BT\nobreakseq{,} United Kingdom}
\author{R.~Salazar} 
\affiliation{University of Texas at Austin\nobreakseq{,}
  Austin\nobreakseq{,} TX 78712\nobreakseq{,}~U.S.A.}
\author{X.~Sarazin} 
\affiliation{LAL, Universit\'{e} Paris-Sud\nobreakseq{,} CNRS/IN2P3\nobreakseq{,} Universit\'{e} Paris-Saclay\nobreakseq{,}
F-91405 Orsay\nobreakseq{,} France}
\author{Yu.~Shitov} 
\affiliation{JINR, 141980 Dubna, Russia}
\affiliation{Imperial College London\nobreakseq{,} London SW7
2AZ\nobreakseq{,} United Kingdom}
\author{L.~Simard} 
\affiliation{LAL, Universit\'{e} Paris-Sud\nobreakseq{,} CNRS/IN2P3\nobreakseq{,} Universit\'{e} Paris-Saclay\nobreakseq{,}
F-91405 Orsay\nobreakseq{,} France}
\affiliation{Institut Universitaire de France\nobreakseq{,} F-75005 Paris\nobreakseq{,} France}
\author{F.~\v{S}imkovic} 
\affiliation{FMFI,~Comenius~Univ.\nobreakseq{,}~SK-842~48~Bratislava\nobreakseq{,}~Slovakia}
\author{A.~Smetana}
\affiliation{Institute of Experimental and Applied Physics\nobreakseq{,} Czech
Technical University in Prague\nobreakseq{,} CZ-12800
Prague\nobreakseq{,} Czech Republic}
\author{K.~Smolek} 
\affiliation{Institute of Experimental and Applied Physics\nobreakseq{,} Czech
Technical University in Prague\nobreakseq{,} CZ-12800
Prague\nobreakseq{,} Czech Republic}
\author{A.~Smolnikov} 
\affiliation{JINR, 141980 Dubna, Russia}
\author{S.~S\"oldner-Rembold}
\affiliation{University of Manchester\nobreakseq{,} Manchester M13 9PL\nobreakseq{,}~United Kingdom}
\author{B.~Soul\'e}
\affiliation{CENBG\nobreakseq{,} Universit\'e de Bordeaux\nobreakseq{,} CNRS/IN2P3\nobreakseq{,} F-33175 Gradignan\nobreakseq{,} France}
\author{I.~\v{S}tekl} 
\affiliation{Institute of Experimental and Applied Physics\nobreakseq{,} Czech Technical University in Prague\nobreakseq{,} CZ-12800 Prague\nobreakseq{,} Czech Republic}
\author{J.~Suhonen} 
\affiliation{Jyv\"askyl\"a University\nobreakseq{,} FIN-40351 Jyv\"askyl\"a\nobreakseq{,} Finland}
\author{C.S.~Sutton} 
\affiliation{MHC\nobreakseq{,} South Hadley\nobreakseq{,} Massachusetts 01075\nobreakseq{,} U.S.A.}
\author{G.~Szklarz}
\affiliation{LAL, Universit\'{e} Paris-Sud\nobreakseq{,} CNRS/IN2P3\nobreakseq{,} Universit\'{e} Paris-Saclay\nobreakseq{,}
F-91405 Orsay\nobreakseq{,} France}
\author{J.~Thomas} 
\affiliation{UCL, London WC1E 6BT\nobreakseq{,} United Kingdom}
\author{V.~Timkin} 
\affiliation{JINR, 141980 Dubna, Russia}
\author{S.~Torre} 
\affiliation{UCL, London WC1E 6BT\nobreakseq{,} United Kingdom}
\author{Vl.I.~Tretyak} 
\affiliation{Institute for Nuclear Research\nobreakseq{,} MSP 03680\nobreakseq{,} Kyiv\nobreakseq{,} Ukraine}
\author{V.I.~Tretyak}
\affiliation{JINR, 141980 Dubna, Russia}
\author{V.I.~Umatov} 
\affiliation{NRC ``Kurchatov Institute", ITEP, 117218 Moscow, Russia}
\author{I.~Vanushin} 
\affiliation{NRC ``Kurchatov Institute", ITEP, 117218 Moscow, Russia}
\author{C.~Vilela} 
\affiliation{UCL, London WC1E 6BT\nobreakseq{,} United Kingdom}
\author{V.~Vorobel} 
\affiliation{Charles University in Prague\nobreakseq{,} Faculty of Mathematics
and Physics\nobreakseq{,} CZ-12116 Prague\nobreakseq{,} Czech Republic}
\author{D.~Waters} 
\affiliation{UCL, London WC1E 6BT\nobreakseq{,} United Kingdom}
\author{A.~\v{Z}ukauskas}
\affiliation{Charles University in Prague\nobreakseq{,} Faculty of Mathematics
and Physics\nobreakseq{,} CZ-12116 Prague\nobreakseq{,} Czech Republic}
\collaboration{NEMO-3 Collaboration}
\noaffiliation

\date{\today}

\begin{abstract}
We present results from a search for neutrinoless double-$\beta$ (\zeronu) decay using \unit[36.6]{g} of the isotope $^{150}$Nd with data corresponding to a live time of \unit[5.25]{y} recorded with the NEMO-3 detector. We construct a complete background model for this isotope, including a measurement of the two-neutrino double-$\beta$ decay half-life of $T^{2\nu}_{1/2}=$~[9.34 $\pm$ 0.22~(stat.)~$^{+0.62}_{-0.60}$~(syst.)]$\times 10^{18}${y} for the ground state transition, which represents the most precise result to date for this isotope. We perform a multivariate analysis to search for \zeronu decays in order to improve the sensitivity and, in the case of observation, disentangle the possible underlying decay mechanisms. As no evidence for \zeronu decay is observed, we derive lower limits on half-lives for several mechanisms involving physics beyond the Standard Model.  The observed lower limit, assuming light Majorana neutrino exchange mediates the decay, is $T^{0\nu}_{1/2} >$~\unit[2.0]{$\times 10^{22}$ y} at the 90\% C.L., corresponding to an upper limit on the effective neutrino mass of \mbb~$<$~\unit[1.6 -- 5.3]{eV}.
\end{abstract}

\maketitle

\section{Introduction}
\label{Sec:Introduction}
The \nemo detector was operated from February 2003 to January 2011 in the Modane Underground Laboratory (LSM) to search for neutrinoless double-$\beta$ (\zeronu) decay~\cite{TDR}. This nuclear decay violates lepton number conservation by two units. Its observation would therefore provide direct evidence for physics beyond the Standard Model (BSM). The experimental signature of $\beta\beta$ decay involves the detection of two simultaneously emitted electrons from a common decay vertex. The \zeronu decay half-life for an isotope with mass number $A$ and atomic number $Z$ is 
\begin{equation}\label{Eq:halflife_zeronu}
 T^{0\nu}_{1/2}(A,Z)^{-1} = g_{A}^{4}G^{0\nu}(Q_{\beta\beta},Z)\vert M^{0\nu}(A,Z)\vert^{2}\xi^{2},
\end{equation}
where $g_{A}$ is the axial vector coupling constant, $G^{0\nu}$ is a phase space factor that depends on $Z$ and the nuclear transition energy $Q_{\beta\beta}$, $M^{0\nu}$ is the nuclear matrix element (NME), and $\xi$ represents a parameter of the underlying BSM physics model. Both $G^{0\nu}$ and $M^{0\nu}$ depend on the assumed BSM decay mechanism. 

Under the assumption that a massive Majorana neutrino mediates the decay, the half-life depends on the effective Majorana neutrino mass \mbb such that $\xi^{2} = (\langle m_{\nu} \rangle / m_{e})^{2}$, where $m_{e}$ is the electron mass~\cite{Rodejohann:2011mu}. Therefore, if \zeronu decay occurs through this mechanism then the measurement of the decay rate could provide information about the absolute mass scale of the neutrino.  
A number of other BSM mechanisms could also mediate a \zeronu decay, such as right handed currents (RHC) or majoron decay modes~\cite{Ali:2007ec,Bamert:1994hb,Carone:1993jv,Mohapatra:2000px}. Data from the \nemo experiment is used to search for evidence of each of these decay mechanisms. 

The isotope $^{150}$Nd is a particularly interesting candidate to search for \zeronu decay due to its large $Z$ and its $Q_{\beta\beta}$ value of \unit[$\lbrack 3371.38 \pm 0.20 \rbrack $]{keV}~\cite{Kolhinen:2010zza}, which together yield the largest $G^{0\nu}$ of all \zeronu decay candidates.  Since $Q_{\beta\beta}$ is large, the region of interest in the distribution of the total energy of both electrons  lies above most natural radioactive backgrounds
for \zeronu decay modes that do not involve the emission of additional neutral particles. Nevertheless, the rate of \zeronu decay is expected to be very low, and a complete understanding of all potential background rates is critical.  This includes the two-neutrino double-$\beta$ decay process (\twonu), as well as other sources of natural radioactivity both in and around the active detector volume that may produce signal-like events.

Two previous experiments have directly measured the \twonu decay half-life of  $^{150}$Nd to the ground state of $^{150}$Sm. A group at the University of California, Irvine conducted the first measurement, using a time projection chamber (TPC) located approximately \unit[72]{m} underground at the Hoover Dam~\cite{Elliott:1993cp}. With \unit[15.5]{g} of Nd$_{2}$O$_{3}$ enriched to $91\%$ of the isotope $^{150}$Nd and a live time of \unit[262]{d}, the half-life was measured to be \unit[$T_{1/2}^{2\nu} = \lbrack 6.75 ^{+0.37}_{-0.42}~\mathrm{(stat)} \pm 0.68~(\mathrm{syst})\rbrack$]{$\times 10^{18}$y}~\cite{PhysRevC.56.2451}. The Institute for Theoretical and Experimental Physics (ITEP) in Moscow conducted
another TPC experiment in collaboration with the Institute for Nuclear Research (INR)~\cite{MoscowTPC2,MoscowTPC}. Using a live time of \unit[90]{d} and \unit[51.5]{g} of Nd$_{2}$O$_{3}$, the ITEP group measured the half-life to be \unit[$T_{1/2}^{2\nu} = \lbrack 18.8^{+6.6}_{-3.9}~\mathrm{(stat)} \pm 0.19~(\mathrm{syst}) \rbrack$]{$\times 10^{18}$y}.  The \nemo Collaboration has previously published results on the \twonu decay half-life using half of the total exposure, corresponding to \unit[927.4]{d} and \unit[36.6]{g} of $^{150}$Nd. The half-life from this data set was measured to be \unit[$T_{1/2}^{2\nu} = \lbrack 9.11 ^{+0.25}_{-0.22}~\mathrm{(stat)} \pm 0.63~(\mathrm{syst}) \rbrack$]{$\times 10^{18}$y}~\cite{Nd150_pub}. 

In this article, we present an updated measurement of the \twonu decay half-life of $^{150}$Nd using the full data set from the \nemo experiment, which includes an additional \unit[991.1]{d} of live time compared 
to Ref.~\cite{Nd150_pub}. We construct a full background model for this isotope and evaluate the sources of systematic uncertainties. Using this background model and a measurement of the \twonu decay rate, we search for \zeronu decays mediated by several possible BSM mechanisms.

In previous publications, limits on \zeronu decay rates have been obtained using only the total energy of the two $\beta$ particles emitted in the decay, as this is the most sensitive single observable for the \zeronu signal.  However, there is additional information embedded in other kinematic and topological observables from the two decay electrons, which can be used to better discriminate the \zeronu signal from backgrounds and to disentangle the underlying BSM mechanisms. The design of the \nemo detector allows for reconstruction of the topology and kinematics of final state particles, and thus provides a unique opportunity to study several observables and their correlations through a multivariate analysis (MVA), thereby improving the sensitivity to \zeronu decays. Here, we present the first \zeronu limits derived from the output distribution of a MVA using a boosted decision tree (BDT)~\cite{TMVA}.

\section{The NEMO-3 Detector}
\label{Sec:Detector}

\begin{figure*}[!t]
    \subfigure{\begin{overpic}[scale=0.4]{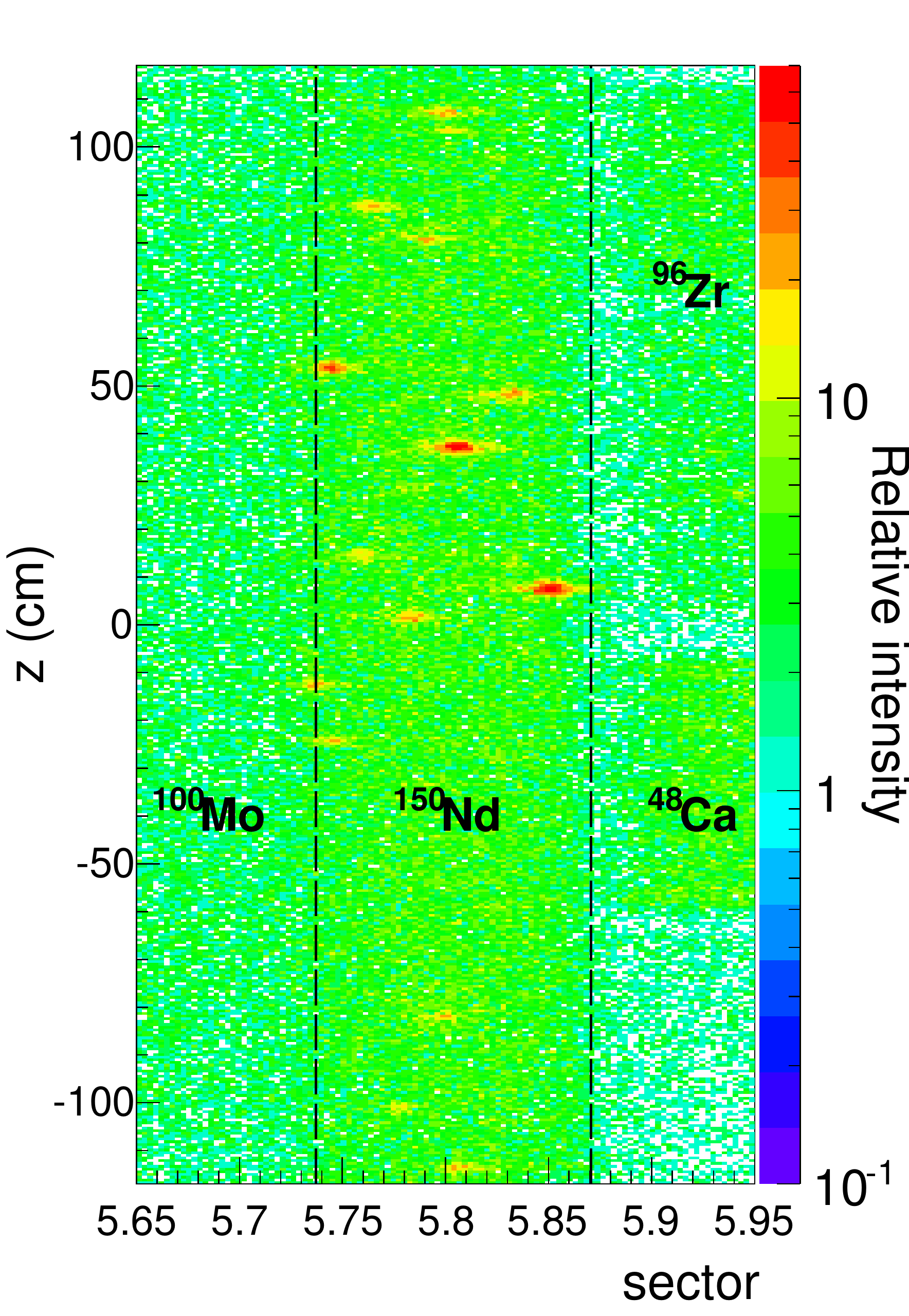}\put(95,0){\large\textbf{(a)}}\end{overpic}\label{Fig:foilMap_HS}}
    \hspace{1.8cm}
    \subfigure{\begin{overpic}[scale=0.4]{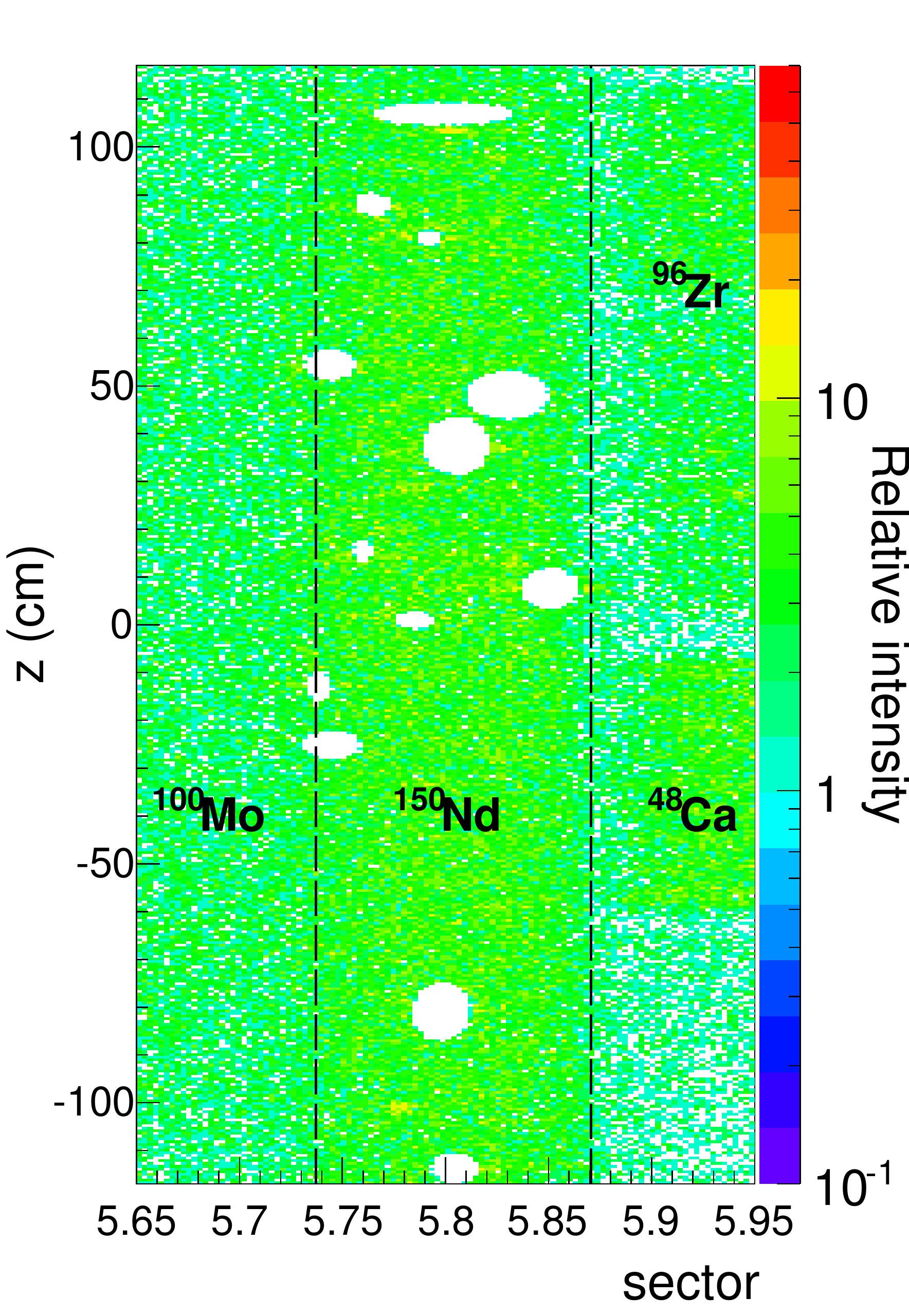}\put(95,0){\large\textbf{(b)}}\end{overpic}\label{Fig:foilMap_noHS}}
    \caption{\label{Fig:foilMap} Two-dimensional representation of the $^{150}$Nd source foil and neighboring isotopes.  These images show the distribution of reconstructed electron track vertices from events that contain one electron and one $\gamma$ ray in the final state both~\subref{Fig:foilMap_HS} before and~\subref{Fig:foilMap_noHS} after the removal of  hot spot regions. The sector number is defined as $(20/2\pi)\cdot \phi$, where $\phi$ is the azimuthal angle.}
\end{figure*}

The \nemo detector, described in detail in Ref.~\cite{TDR} and more recently in Ref.~\cite{Arnold:2015wpy}, is designed to search for \zeronu decays with seven different isotopes simultaneously by reconstructing the full topology of each decay.  The detector is cylindrical in shape with a height of \unit[3]{m} and diameter of \unit[5]{m}, and is divided into 20 equal sectors. Thin source foils with a thickness of \unit[40--60]{mg/cm$^{2}$} are constructed from the different $\beta\beta$-decay isotopes and are distributed around the different sectors at a fixed radius of approximately \unit[155]{cm}.  

The source foils are located between two concentric tracking chambers, commonly referred to as the inner and outer chambers, composed of 6180 drift cells operating in Geiger mode. The Geiger cells in each chamber are strung vertically and are arranged radially in a 4-2-3 layer configuration where the four-cell-wide layer is closest to the source foil, and the three-cell-wide layer is closest to the calorimeter (see Ref.~\cite{TDR}). The cells operate in a gas mixture composed primarily of helium with 4\% ethanol, 1\% argon and 0.15\% water vapor. The tracking detector provides three-dimensional measurements of trajectories and decay vertices of charged particles.  
The vertex resolution is determined by comparing the reconstructed positions of calibration sources to their known locations~\cite{TDR,Arnold:2015wpy}.  The resolution depends on both the electron energy and the longitudinal position of the decay vertex. On average, the transverse ($xy$ coordinate) and longitudinal ($z$ coordinate) resolutions are \unit[0.5]{cm} and \unit[0.8]{cm}, respectively, for \unit[1]{MeV} electrons~\cite{TDR}.

Surrounding the tracking chambers on all sides are calorimeter walls composed of 1940 scintillator blocks coupled to low-radioactivity 3-inch and 5-inch photomultiplier tubes (PMT), which provide both timing and energy measurements. The scintillator blocks on the top and bottom of the detector located in the gaps between the Geiger cell layers are referred to as the ``petal blocks''. The energy and timing resolutions are determined using calibration sources as described in~\cite{TDR,Arnold:2015wpy}. The energy resolution of the calorimeter is (14.1--17.7)\%  (FWHM) for the optical modules with 3-inch and 5-inch PMTs, respectively, and  
the average time resolution is $\approx 250$~ps for electrons with an energy of \unit[1]{MeV}~\cite{TDR}.

To ensure the correct calibration of optical modules in the run periods between absolute energy calibration runs, which occur every 2--3 weeks, a laser survey was conducted twice daily~\cite{TDR}.  This laser survey identifies PMTs whose gains have shifted significantly between absolute calibration runs and also provides corrections to time measurements from the PMTs. The counting rate for each optical module is also monitored over the course of the experiment. Any PMT with large gain variations or irregular counting rates is excluded from this analysis for the periods during which they are identified as unstable. The systematic uncertainty on the energy calibration is 
$\approx 1\%$, which is estimated from the average uncertainty on the gain for all of the remaining PMTs used in the analysis.

Surrounding the calorimeter is a solenoid coil that produces a magnetic field of \unit[25]{G} parallel to the axis of the cylinder.  This magnetic field provides discrimination between electrons and positrons through their track curvature. 

The overburden at LSM is approximately \unit[4800]{m.w.e.}, which significantly reduces the cosmic ray flux. The decays of naturally occurring $^{238}$U, $^{232}$Th, and their daughter decay products located in the rock surrounding the LSM laboratory can produce external neutron and $\gamma$-ray fluxes that constitute backgrounds to the $\beta\beta$-decay signal~\cite{bkg_paper}.  Several components of passive shielding around the \nemo detector reduce this external background flux. 

After the first year of operation, an anti-radon facility was installed to reduce the amount of $^{222}$Rn diffusion into the detector. The facility comprises an airtight tent to provide a buffer zone between the laboratory air and the active detector volume, and a carbon trap air filtration system that removes residual $^{222}$Rn in this buffer zone. The use of this facility reduces the overall activity of $^{222}$Rn decay products in the active detector by a factor of six~\cite{bkg_paper}. 

The $^{150}$Nd source foil is located in Sector 5 of the detector between a strip of $^{100}$Mo and strips containing $^{96}$Zr and $^{48}$Ca as shown in Fig.~\ref{Fig:foilMap}. The foil is made from a Nd$_{2}$O$_{3}$ powder provided by the INR. It is enriched to $( 91.0 \pm 0.5 )\%$ of the isotope $^{150}$Nd using electromagnetic separation and chemically purified. The Nd$_{2}$O$_{3}$ powder was mixed with a concentration of $8\%$ polyvinyl alcohol for bonding to two Mylar backing films to produce a composite source foil with a total mass of \unit[56.68]{g} for insertion into the \nemo detector.  The total mass of $^{150}$Nd in the foil is \unit[$(36.6 \pm 0.2 )$]{g}~\cite{TDR}. 

Two data taking phases are defined for \nemo detector operation. They correspond to the run periods before (Phase 1) and after (Phase 2) the installation of the anti-radon facility. The combined data from both phases yield a live time of \unit[5.25]{y}, which corresponds to an exposure for $^{150}$Nd of \unit[0.19]{kg$\cdot$y}. This live time is larger than the one used in~Ref.~\cite{Arnold:2015wpy} due to a slightly looser selection of run conditions in this analysis.

\section{Reconstruction and event selection}
\label{Sec:Reconstruction}

A \nemo event trigger requires an energy deposit in the calorimeter of approximately \unit[$150$]{keV} in combination with a series of Geiger hits in three different layers occurring within \unit[$6.14$]{$\mu$s} in the same or a nearby sector~\cite{TDR}. A tracking algorithm constructs helical tracks from these prompt Geiger hits.  The tracks are extrapolated to the source foil radius to obtain a vertex location and to the calorimeter wall to associate an optical module for energy and time measurements. 

An electron is defined as a track that is associated with an isolated optical module on the calorimeter wall. Several criteria are applied to select electrons with good quality track and energy reconstruction. The curvature of the track is required to be consistent with a negatively charged particle as determined by the magnetic field, where the charge is determined by assuming the track originates at the source foil. The reconstructed track length is required to be at least \unit[50]{cm} for events with a single track to ensure that the vertex and calorimeter impact points are accurately extrapolated.  To further improve the track extrapolation to the vertex, the tracks must have an associated Geiger cell hit within one of the first two layers of the tracker closest to the source foil. In addition, no more than one prompt Geiger hit that is unassociated to the track is allowed within a distance of \unit[15]{cm} from the vertex. 

For events with two tracks, the track length requirement is relaxed to \unit[$\geq$~30]{cm}. The separation between each individually reconstructed vertex is required to be \unit[$\Delta R = \sqrt{(\Delta x)^{2} + (\Delta y)^{2}} \leq 4$]{cm} (transverse direction) and \unit[$\Delta z \leq 8$]{cm} (longitudinal direction) to ensure that the two tracks are associated to a common event. This separation is significantly larger than the vertex resolution. Using the data and predicated background rates, we estimate the background from two simultaneous single $\beta$ decays in this allowed region to be negligible.

To ensure optimal energy reconstruction, tracks must enter the calorimeter blocks from the front face. Only events with electrons of energies \unit[$E_{e} \geq 300$]{keV} are considered in this analysis, as the rate of background decays rapidly increases below this threshold. Electrons from $^{150}$Nd $\beta\beta$ decays achieve much higher energies due to the large $Q_{\beta\beta}$ value. Electron tracks extrapolated to the petal blocks closest to the source foil are rejected in this analysis because of the poorer energy and time resolution for these events~\cite{TDR,Arnold:2015wpy}. 

Geiger hits that are recorded in a time window \unit[6.14--700]{$\mu$s} after the calorimeter trigger are also stored. These delayed hits are used to identify $\alpha$ particles from $^{214}$Bi--$^{214}$Po cascades. The rate of these cascades is used to constrain the activity of $^{214}$Bi, which is an important background in the search for \zeronu decays.  A $^{214}$Bi--$^{214}$Po cascade is characterised by a ($\beta$,$\gamma$) decay followed by an $\alpha$ decay with a delay consistent with the $^{214}$Po half-life of $T_{1/2}$ = \unit[164.3]{$\mu$s}~\cite{214_scheme}. The trajectory of an $\alpha$ particle
with the same energy as an electron will not be significantly affected by the magnetic field because of its larger mass. Due to large ionization energy losses, the lengths of tracks produced by $\alpha$ particles from radioactive decays are shorter than~\unit[40]{cm}~\cite{bkg_paper}. Therefore, $\alpha$ particles are identified in the \nemo detector as short, straight tracks composed of delayed Geiger hits.

Many decays that are sources of background are accompanied by the emission of high energy $\gamma$ rays. A $\gamma$ ray is identified as either a single calorimeter hit or a cluster of neighboring hits, with no tracks attributed to its position and an energy of \unit[$E_{\gamma} \geq 200$]{keV}. No more than one prompt Geiger hit located within \unit[15]{cm} of any $\gamma$-ray hits is allowed. This reduces the misidentification of electrons as $\gamma$ rays, while allowing for low level tracker noise which is not simulated.  Calorimeter noise can be characterized as isolated calorimeter hits with \unit[$E_{\gamma} < 150$]{keV} and no track association. 
Since calorimeter noise is also not simulated for this analysis, calorimeter hits meeting these criteria are ignored in the event selection process.

Vertices are required to be located within the boundaries of the $^{150}$Nd source foil, defined as \unit[$\vert\mathrm{z}\vert \leq 117$]{cm} and $5.7371\leq\mathrm{sector}\leq 5.8706$ (see Fig.~\ref{Fig:foilMap}). Several regions of high activity are identified in the $^{150}$Nd source foil that correspond to localized contamination from $^{234m}$Pa and $^{207}$Bi. These regions are referred to as hot spots and are evident in the vertex distribution shown in Fig.~\ref{Fig:foilMap_HS}. The origin of these hotspots in unknown, though they were likely introduced during the foil production or installation.  Due to the ability to accurately reconstruct track vertices, these hot spots are easily identified and removed in the event selection process for this analysis. Event vertices are binned in Fig.~\ref{Fig:foilMap_HS} according to the resolution of the \nemo tracker~\cite{TDR}. The distribution of the number of events in each bin is fitted with a Poisson function, and bins with event rates greater than 3$\sigma$ fluctuation from the average are identified as potential hot spot candidates. Clusters of hot spot candidate bins are fitted with a 2-dimensional Gaussian function, where the mean and width of each fit is used to define an ellipse containing a hot spot region. The distribution of event vertices after removing these hot spot regions is shown in Fig.~\ref{Fig:foilMap_noHS}. This corresponds to a loss of an area of $\approx 105.9$~cm$^{2}$, which is $\approx 7\%$ of the $^{150}$Nd foil area. A uniform distribution of signal and background isotopes is assumed for the remaining foil area. 

Discrimination between decays that are internal and external to the source foil is achieved with two time-of-flight (TOF) probabilities that are defined using the energy and time measurements from the calorimeter and the distances travelled by each particle in the event~\cite{TDR,Arnold:2015wpy}.  A $\chi^{2}$ formula is constructed by comparing the difference between the measured times and the expected time of flight calculated from each particle's trajectory and the timing of calorimeter signal. Energy losses in the tracker volume and the uncertainties on all quantities are taken into account in the $\chi^{2}$ calculation.  Since $\gamma$ rays do not leave hits in the tracker, their trajectories are calculated as a straight line from the geometrical center of the front face of the first calorimeter block in a cluster to the reconstructed event vertex, which is given by the electron track intersection with the foil. A $\chi^{2}$ formula is derived for both an internal decay hypothesis, which assumes a common decay vertex in the source foil, and an external decay hypothesis that the initial decay vertex is external to the source foil. Probability distributions are formed from both $\chi^{2}$ values, and are used as part of the event selection criteria to select internal or external-like decay topologies. Details of the internal ($P_{\mathrm{int}}$) and external ($P_{\mathrm{ext}}$) TOF probability calculations are provided in Refs.~\cite{TDR} and~\cite{Arnold:2015wpy}.

\section{Analysis technique}
\label{Sec:Analysis_technique}

The \nemo analysis technique involves comparisons of experimental data to Monte Carlo (MC) simulations of radioactive decays. The event generator DECAY0~\cite{DECAY4} provides final-state particles with kinematics following the decay schemes of all isotopes considered in this analysis.  The transport of these particles is simulated using a complete description of the \nemo detector in GEANT-3.21~\cite{Brun:1987ma}, which provides modelling of particle interactions in the detector material. 

Events are simulated and reconstructed under the same detector conditions as the real data with the exception of noise, which is not implemented in the simulation. All data samples are divided into signal and background channels based on their event topologies and final state particle content. The signal channel for detection of $\beta\beta$ decays consists of events with only two electrons in the final state. The background channels are divided into three categories depending on the origin of the decay: internal or external to the source foil, and decays from radon progeny in the tracking chambers. These categories can be further subdivided into more specific decay channels based on the particle content of the decay as will be discussed in Sec.~\ref{Sec:Background_model}. The decay rates of all isotopes are measured \emph{in situ} by fitting the MC distributions to the \nemo data using a Poisson log-likelihood function on binned observables from each channel that provide the optimal sensitivity to the isotopes of interest. We minimize the expression
\begin{equation}
\label{Eq:LogLik}
\begin{split}
-2\mathrm{ln}(L)& = -2\sum\limits_{i,n}\left[ \right.-(s_{i,n}+\sum\limits_{j} b_{i,j,n}) + \\
 & \quad d_{i,n}\mathrm{ln}\left ( s_{i,n}+\sum\limits_{j}b_{i,j,n} \right) - \mathrm{ln}(d_{i,n}!) \left. \right],
\end{split}
\end{equation}
where the sum is performed over all bins $i$ of the observable from each channel $n$ used in the fit. The minimum of $-2\mathrm{ln}(L)$ is obtained with the number of signal events, $s$, and events from each background, $b_{j}$, that best describe the observed data, $d$, in each channel. If the activity of a radionuclide is known with high precision through independent measurements (e.g. high purity germanium detectors) or if the activity measured in a particular decay channel is considered more robust due to the control of systematic uncertainties then Gaussian constraints are introduced into the likelihood. The center and width of the constraints are determined from the best fit activities and uncertainties in these more accurate measurements. 

Many of the radionuclides considered in this analysis produce events in multiple decay channels with varying degrees of sensitivity and correlations with other components of the background model.  Therefore, preliminary estimates of background rates are obtained by fitting each background channel distribution separately. In terms of Eq.~\ref{Eq:LogLik}, this means that the likelihood is minimized for a single channel $n$, and the signal $s$ represents the isotope(s) of interest for that channel. The final measurements of each background isotope activity and the $^{150}$Nd \twonu decay rate result from a simultaneous likelihood fit to the selected observables for the signal and all backgrounds (S+B) according to Eq.~\ref{Eq:LogLik}, which directly accounts for correlations between isotopes displaying similar kinematics. The normalization of all MC distributions reflects the results from this global S+B fit unless otherwise stated.

\section{The background model}
\label{Sec:Background_model}

Decays due to internal backgrounds from source foil impurities can mimic the $\beta\beta$-decay signal through several different mechanisms such as a single-$\beta$ decay to the ground state of the daughter nucleus combined with M\o ller scattering, or $\beta$ decay to an excited state of the daughter nucleus followed by the emission of a conversion electron or $\gamma$ ray that undergoes Compton scattering in the foil. The decay of $^{214}$Bi ($^{222}$Rn progeny) near the source foil can produce signal-like events in an analogous manner to internal background decays. Decays from external backgrounds typically result in signal-like events through decays that yield high energy $\gamma$ rays with subsequent pair production or undergo double Compton scattering in the source foil. This section describes each of the decay channels used to estimate the background rates from all of the decay sources considered in this analysis.  The activities and corresponding numbers of events from each background process contributing to the two-electron signal decay channel are presented in Sec.~\ref{Sec:2vBB_measurement}.

\subsection{External backgrounds\label{Sec:External_backgrounds}}

Two channels are used to measure the external background rates.  The single-electron crossing channel (SEC) is used to select events where a single electron travels across the detector.  These events are identified by two tracks with opposite curvature associated with scintillator hits that are typically located on opposite sides of the source foil. This selects a single particle because curvature is reconstructed without timing information and assumes an outgoing particle momentum with respect to the source foils. The timing of the calorimeter hits is required to be consistent with the external TOF hypothesis, i.e., $P_{\mathrm{ext}} \geq$~4\% and $P_{\mathrm{int}} \leq$~1\%. The energy of the (temporally) first calorimeter hit is required to be \unit[$E_{e} \geq$ 0.2]{MeV} and for the second calorimeter hit \unit[$E_{e} \geq$ 0.3]{MeV}, corresponding to a total energy of \unit[$E_{\mathrm{tot}} \geq$ 0.5]{MeV}. Externally produced electrons with energies below this minimum are not of concern as background to the $\beta\beta$-decay measurement due to the large $Q_{\beta\beta}$ value of $^{150}$Nd.

\begin{figure}[t]
    \subfigure {\begin{overpic}[scale=0.4]{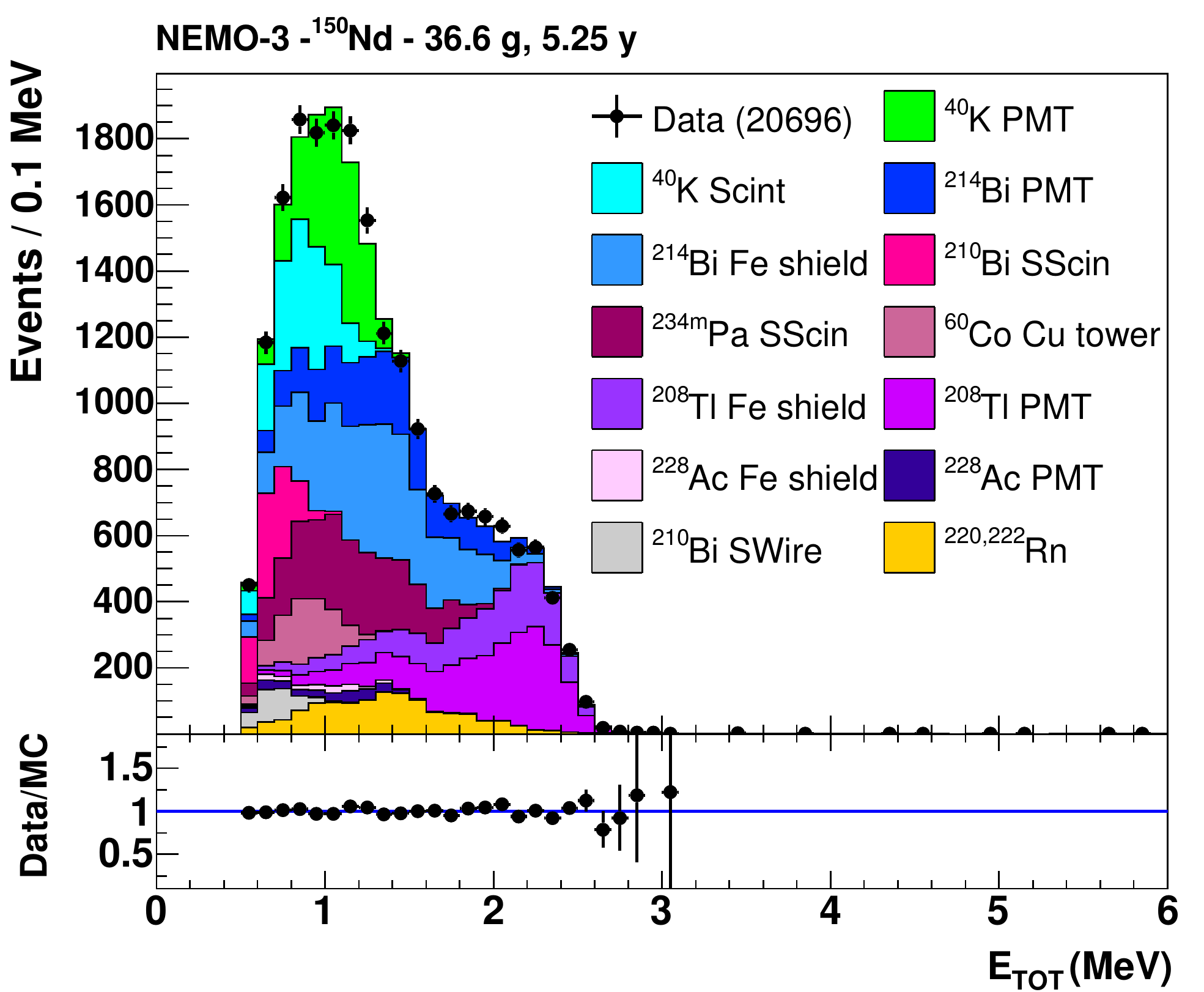}\put(200,60){\large\textbf{(a)}}\end{overpic}\label{Fig:OCE_totalE}}\\
    \subfigure{\begin{overpic}[scale=0.4]{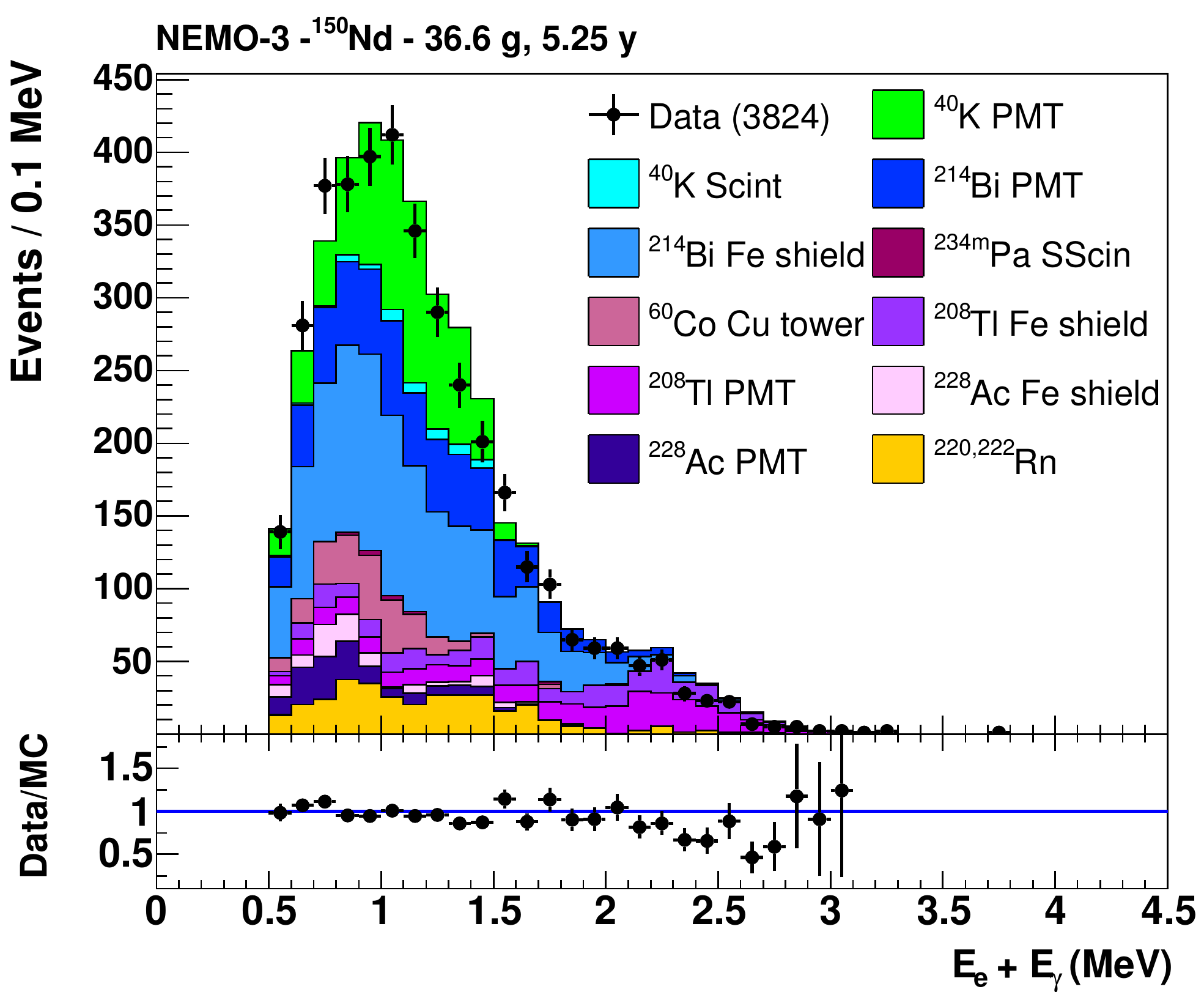}\put(200,60){\large\textbf{(b)}}\end{overpic}\label{Fig:1e1gExt_totalE}}    
    \caption{\label{Fig:External_distributions} Distributions of the total energy for the \subref{Fig:OCE_totalE} SEC and \subref{Fig:1e1gExt_totalE} 1e1$\gamma$-Ext background channels. The data are compared to the MC simulation.}
\end{figure}

Backgrounds from several radionuclides located in the PMTs, scintillator surfaces (SScin), inside the scintillators (Scint), iron shielding (Fe shield), internal copper tower (Cu tower), radon ($^{222}$Rn) and thoron ($^{220}$Rn) progenies in the air, wire surfaces (SWire), and the external neutron flux are all considered in the external background model for the $^{150}$Nd source foil.  The external background model is an effective model that does not contain an exhaustive list of all possible external sources of background. We omit isotopes with distributions that are strongly correlated with other components since this simplification has no impact on the final result. The purpose of this procedure is only to provide an accurate description of the total external $\gamma$ flux~\cite{bkg_paper}. The $E_{\mathrm{tot}}$ distribution is shown in Fig.~\ref{Fig:OCE_totalE} for events meeting the SEC channel selection criteria. This observable provides the maximal separation between the different background contributions in this channel.

The second channel used to measure external background rates is the external gamma-electron (1e1$\gamma$-Ext) channel, where a high energy $\gamma$ ray interacts in the calorimeter and then undergoes Compton scattering in the source foil to produce an electron.  Such events are defined by a single electron accompanied by a $\gamma$ ray, where each particle is identified using the criteria presented in Sec.~\ref{Sec:Reconstruction}. The timing and trajectories of the particles are required to be consistent with the external TOF hypothesis, i.e.,~$P_{\mathrm{ext}} \geq$~4\% and $P_{\mathrm{int}} \leq$~1\%.  The summed energy of the electron and $\gamma$ ray, shown in Fig.~\ref{Fig:1e1gExt_totalE}, is the best observable in this channel for discriminating between the different radionuclides contributing to the external background model. 

The total energy distributions from both the SEC and 1e1$\gamma$-Ext channels are divided into Phases 1 and 2 and fitted simultaneously to provide initial estimates of all external background activities. The division into the separate phases allows for a different normalization of $^{220}$Rn and $^{222}$Rn progenies before and after the installation of the anti-radon facility. The activities of external backgrounds from the same radionuclide in different locations are strongly anti-correlated, as their energy spectra are nearly degenerate. The largest deviations between the results from this fit, which measures the external background rates near the $^{150}$Nd foil, and the results from Ref.~\cite{bkg_paper}, which correspond to the average activities measured across the entire detector using a slightly different effective model, are $[+31, -23]\%$. This is taken as the systematic uncertainty on the external background normalization in the signal channel (see Sec.~\ref{Sec:2vBB_measurement}).  It has a $[+1.2,-0.8]\%$ effect on the \twonu decay half-life, which is small compared to the other systematic uncertainties considered. 

The neutron flux produces a small number of events observed in the SEC channel in the energy range 
$E_{\rm tot}=3$--$8$~MeV. Neutrons do not contribute significantly to any other decay channel.  This demonstrates the effectiveness of the passive shielding described in Sec.~\ref{Sec:Detector}, as well as the excellent background rejection made possible by the timing resolution of the calorimeter.

\subsection{Radon\label{Sec:Radon}}  
  
Radon ($^{222}$Rn) and thoron ($^{220}$Rn) are highly diffusive gases that emanate from rocks in the surrounding environment and may enter the \nemo detector. Both isotopes are also present in various detector materials, and can therefore emanate directly into the active detector volume. Since radon and thoron are both $\alpha$-decay isotopes, these background decays will not trigger the \nemo detector readout as an electron is needed (see Sec.~\ref{Sec:Reconstruction}). However, the $^{214}$Bi and $^{208}$Tl progenies in the radon and thoron decay chains, respectively, can produce decays that mimic the $\beta\beta$ signal. The short half-life of thoron, $T_{1/2} =$~\unit[55.6]{s}, results in low levels of this isotope and its daughters leaking into the \nemo tracker from the outside environment~\cite{bkg_paper}. The contributions from $^{208}$Tl on the surfaces of tracker wires and the source foil are thus found to be negligible in all background channels in this analysis.  

Conversely, the half-life of $^{222}$Rn, $T_{1/2} =$~\unit[3.824]{d}, allows this gas to diffuse into the active detector volume.  The decay products of $^{222}$Rn are mostly positive ions, which deposit themselves on the surfaces of the tracker cathode wires and the surfaces of the source foils.  The contamination of $^{222}$Rn in various locations is measured by selecting a 1e1$\alpha$ decay topology. 

Events in the 1e1$\alpha$ channel are required to have a single electron and at least one delayed $\alpha$ candidate (see Sec.~\ref{Sec:Reconstruction}). An $\alpha$ candidate comprises between 2 and 13 delayed Geiger hits recorded at least \unit[4]{$\mu$s} after the prompt electron hits to prevent misidentification of Geiger cell re-firings as $\alpha$ candidates.  

The $\alpha$ track, which is reconstructed by a straight line fit to the delayed hits, must intersect the foil with a separation of $< 5$~cm in the $xy$ and $z$ coordinates from the electron vertex if it contains only two delayed hits, and $< 10$~cm if it contains more than two delayed hits. The $\alpha$ track must also have an associated hit in one of the first two Geiger cell layers and a total length $\leq$~\unit[40]{cm}. 
  
\begin{figure}[!t]
    \subfigure {\begin{overpic}[scale=0.4]{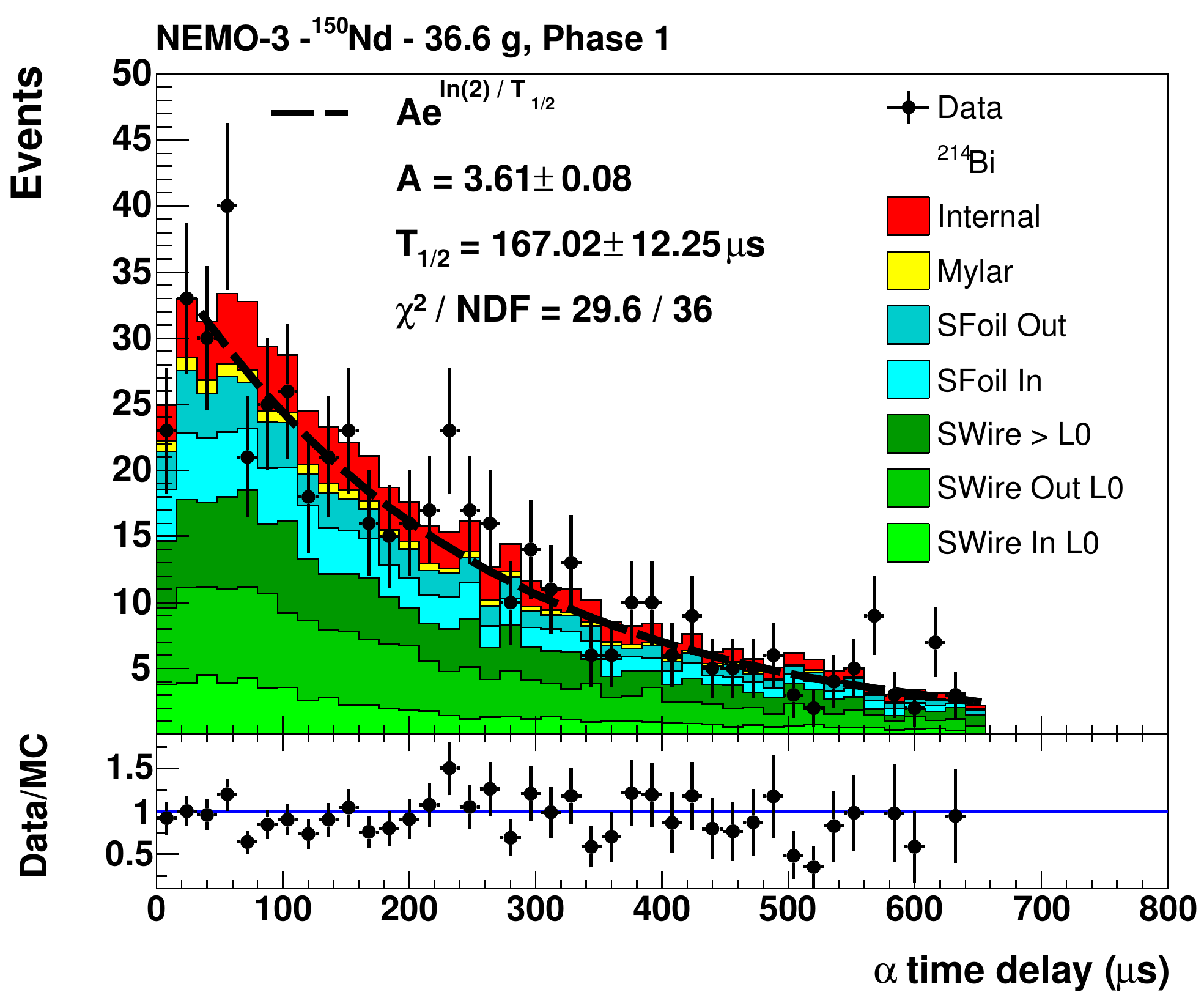}\put(200,60){\large\textbf{(a)}}\end{overpic}\label{Fig:1e1a_alphaTime_P1}}\\
    \subfigure {\begin{overpic}[scale=0.4]{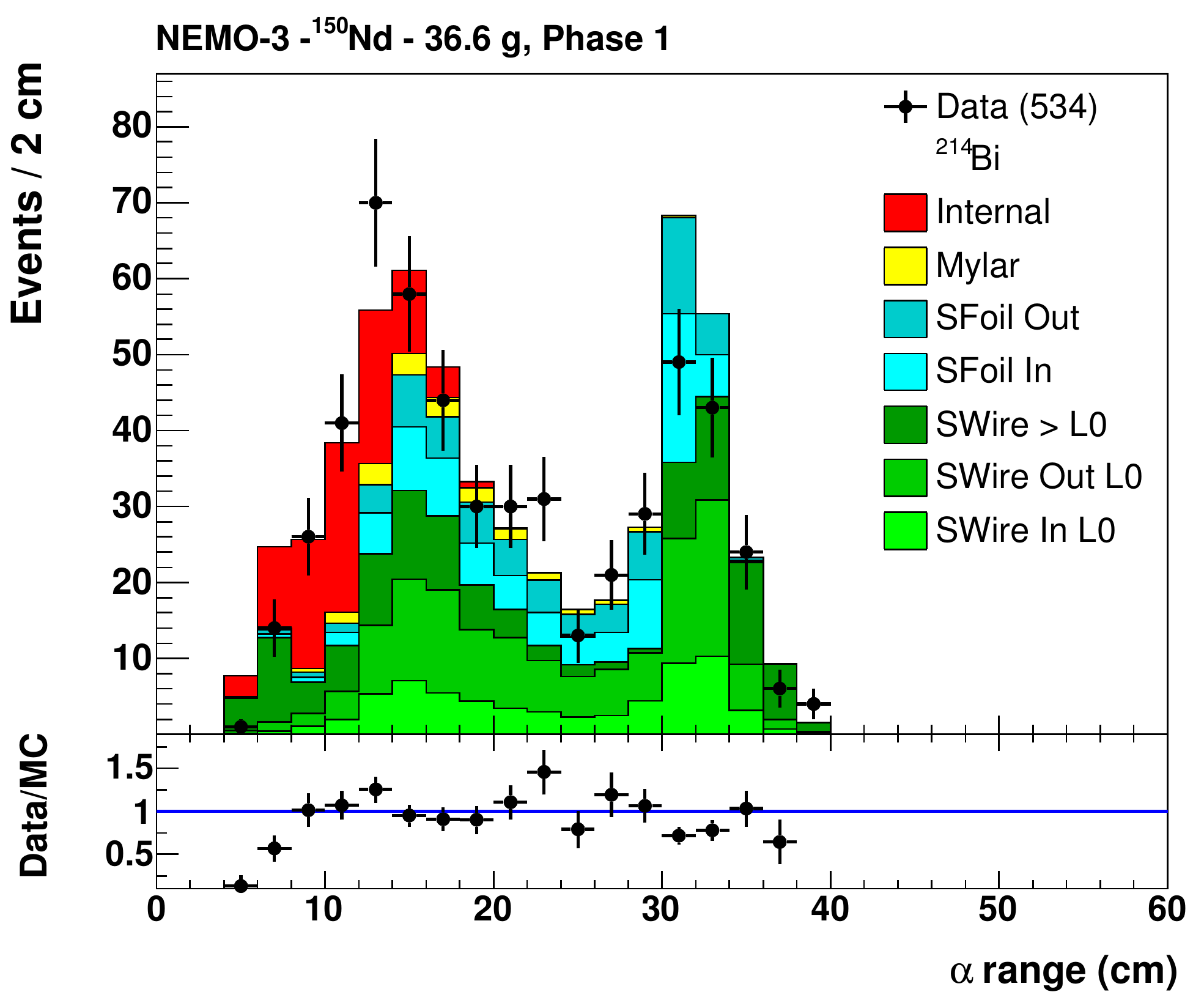}\put(200,60){\large\textbf{(b)}}\end{overpic}\label{Fig:1e1a_alphaLength_P1}}
    \caption{\label{Fig:1e1a_distributions} \subref{Fig:1e1a_alphaTime_P1} Distribution of the average time delay between the event trigger and the $\alpha$ particle detection. Data are fitted with an exponential function, resulting in a $T_{1/2}$ consistent with $^{214}$Po decays. The goodness of fit is demonstrated by the $\chi^{2}$ divided by the number of degrees of freedom (NDF).  \subref{Fig:1e1a_alphaLength_P1} The $\alpha$-range distribution is used in the fit to discriminate between the various $^{214}$Bi components of the background model.  Both distributions are from events that meet the 1e1$\alpha$ channel selection criteria in Phase 1 (high radon period).}
\end{figure}
  
These selection criteria provide a pure sample of $^{214}$Bi events.  This is confirmed by the average delay time of selected $\alpha$ candidates, which is shown in Fig.~\ref{Fig:1e1a_alphaTime_P1} for Phase 1 (high radon period). Fitting the data in Fig.~\ref{Fig:1e1a_alphaTime_P1} with an exponential function results in a decay constant consistent with the known $^{214}$Po half-life. In addition, there is good agreement between data and simulated decays for small $\alpha$ time delays, indicating that the event selection properly suppresses correlated Geiger cell noise. 

The MC samples in Fig.~\ref{Fig:1e1a_distributions} correspond to $^{214}$Bi decays simulated in several different locations. The $\alpha$ range provides the best separation between these various $^{214}$Bi contributions as shown in Fig.~\ref{Fig:1e1a_alphaLength_P1}.  Since $\alpha$ particles are highly ionizing, those emitted from decays internal to the source foil and in the Mylar backing film have short ranges.  In contrast, the $\alpha$ particles from decays on the surface of the foil (SFoil) and the tracker wires (SWire) can travel longer distances as a result of the lower density of the tracker gas mixture relative to the source foil. The two peaks in the $\alpha$-range distribution shown in Fig.~\ref{Fig:1e1a_alphaLength_P1} are due to the gap between the first four layers of Geiger cells and the second group of Geiger cell layers.  Since a hit is required in one of the first two Geiger cell layers from both the electron and $\alpha$ tracks, these selection criteria result in larger sensitivity to $^{214}$Bi decays from the Geiger layer closest to the source foil, referred to as L0, compared to events coming from Geiger layers further away from the foil. 

The large ionization energy losses experienced by $\alpha$ particles if they travel through the source foil prevents them from crossing the foil completely.
The contributions to the $^{214}$Bi activities on the inner (In) and outer (Out) parts of the tracker are therefore estimated separately by dividing the event sample based on the $\alpha$ track location with respect to the source foil and fitting the $\alpha$ range distribution.  The activities of the $^{214}$Bi contamination internal to the foil and in the Mylar backing film are not expected to decrease over the lifetime of the experiment. Therefore, the $\alpha$ range is fitted with a single parameter for each of these components over Phases 1 and 2.  The SWire and SFoil components are allowed to vary between Phases 1 and 2 to account for the lower $^{214}$Bi contamination in the tracker after the installation of the anti-radon facility. 

There is some disagreement between the simulation and data at very short $\alpha$ ranges, where many events originate from inside the source foil, which is likely caused by mis-modelling of $\alpha$-particle energy loss in the source foil. A $23\%$ systematic uncertainty on the internal $^{214}$Bi activity is estimated from the difference in activities measured between the 1e1$\alpha$ and other decay channels used to constrain this decay rate as will be discussed in the following sections. In addition, there is a $10\%$ systematic uncertainty on the SFoil and SWire components of the $^{214}$Bi decay rates which is estimated in the same way. The $\alpha$ range distribution of data and the MC simulation are in agreement within these systematic uncertainties.

\subsection{Internal backgrounds\label{Sec:Internal_bkgs}}

The $^{150}$Nd source foil contamination was measured with high purity germanium (HPGe) detectors prior to insertion into the \nemo detector and after decommissioning the detector, which revealed a number of background radionuclides internal to the foil~\cite{TDR}. Three background channels are defined (1e, 1e1$\gamma$ and 1e2$\gamma$) to measure these internal background decay rates \emph{in situ} and compare them to the HPGe measurements.

\subsubsection{1e channel \label{Sec:1e_channel}}

We use the single-electron (1e) channel to measure the activity of internal $^{40}$K and $^{234m}$Pa, and of $^{210}$Bi on the surfaces of wires and the foil. These radionuclides primarily undergo $\beta$ decay to the ground state of their daughter radionuclides. Events in the 1e channel contain only a single electron track meeting the selection criteria in Sec.~\ref{Sec:Reconstruction}.  No delayed Geiger hits indicative of $\alpha$ particles are allowed within \unit[15]{cm} of the vertex to suppress the radon contribution.

\begin{figure}[htbp]
   {\includegraphics[scale=0.4]{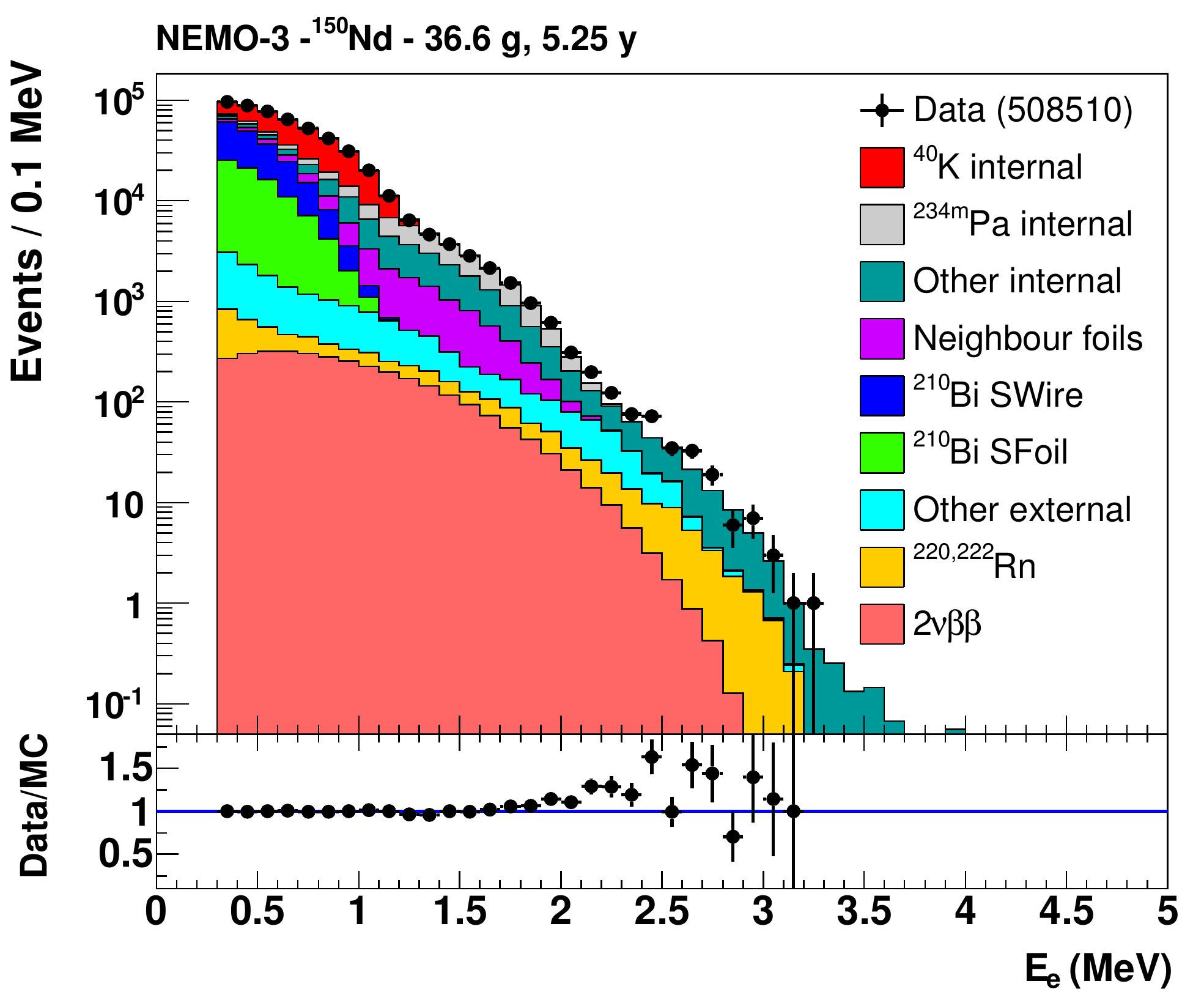}}%
    \caption{\label{Fig:1e_distribution}  Distribution of the electron energy ($E_e$)  in the one-electron channel for data and the background model. The $^{210}$Bi SWire and SFoil components are the dominant contributions to the external background.}
\end{figure}

External backgrounds contribute a large number of events in this channel because internal and external probabilities cannot be calculated with a single particle.  Their rates are constrained by the SEC and 1e1$\gamma$-Ext channels described previously. 
Events from neighboring source foils ($^{100}$Mo, $^{96}$Zr, and $^{48}$Ca) are sometimes reconstructed within the $^{150}$Nd source foil boundaries due to the vertex resolution of the tracker (see Sec~\ref{Sec:Detector}). In the background model, we fix event rates from these neighboring foils to values measured in dedicated analyses~\cite{Zr96_thesis,Ca48_paper,Arnold:2015wpy}. Finally, the \twonu decay of $^{150}$Nd also contributes to the 1e channel as shown in Fig.~\ref{Fig:1e_distribution} if one of the electrons is not properly reconstructed.

The electron energy distribution shown in Fig.~\ref{Fig:1e_distribution} is used to estimate the activities of $^{40}$K, $^{234m}$Pa, and $^{210}$Bi. The $^{210}$Bi SFoil and SWire components are highly correlated ($-80\%$) due to the similarity of their energy spectra. Events from $^{40}$K decays are concentrated at low energies, resulting in a $-30\%$ correlation between this activity and most of the external backgrounds, and a $-80\%$ correlation with the $^{210}$Bi SFoil component. The normalization of $^{234m}$Pa events, which contribute to the full energy spectrum, is largely uncorrelated with the external backgrounds, and exhibits a $-50\%$ correlation with $^{40}$K and $^{214}$Bi.

We observe good agreement between the simulation and data.  There is a slight excess of data for energies above $\approx$ \unit[2]{MeV}, which is attributed to $^{214}$Bi decays. Since the energy spectra of the different $^{214}$Bi contributions in this channel are degenerate, it is not possible to determine which component (internal or external) may be responsible for the upward fluctuation. The event rates for all $^{214}$Bi components are constrained to the average of the activities measured in the 1e1$\alpha$ and 1e1$\gamma$ channels, which provide better sensitivity to this isotope. A $23(10)\%$ systematic uncertainty on the normalization of the internal(external) $^{214}$Bi contribution is estimated from these more sensitive decay channels. These decays form the dominant contribution above \unit[2]{MeV} to the ``Other internal'' stacked histogram in Fig.~\ref{Fig:1e_distribution}. $^{214}$Bi events coming from radon decays in the tracker are denoted in this Fig. as $^{220,222}$Rn. The total uncertainty on the expected event rates from both of these $^{214}$Bi contributions accounts for most of the excess at \unit[$E_e>2$]{MeV} in Fig.~\ref{Fig:1e_distribution}. 
 
\subsubsection{1e(N)$\gamma$ channels}
 
The most critical background for the \zeronu search with the $^{150}$Nd source foil is the isotope $^{208}$Tl.  It primarily undergoes $\beta$ decay to the excited states of $^{208}$Pb, which subsequently decay to the ground state through the emission of high-energy $\gamma$ rays with energies up to \unit[$E_\gamma\approx 2.6$]{MeV}. Thus, the 1e1$\gamma$ and 1e2$\gamma$ channels are used to measure the decay rate of the $^{208}$Tl background as they provide the best sensitivity. The activities of $^{207, 214}$Bi and $^{152,154}$Eu are also measured. We assume that $^{228}$Ac and $^{212}$Bi, which also contribute to the total event rate in these channels, are in secular equilibrium with $^{208}$Tl.  We therefore fit these activities with a single parameter. The same is done for the activities of $^{214}$Pb and $^{214}$Bi. 

\begin{figure}[t!]
    \subfigure {\begin{overpic}[scale=0.4]{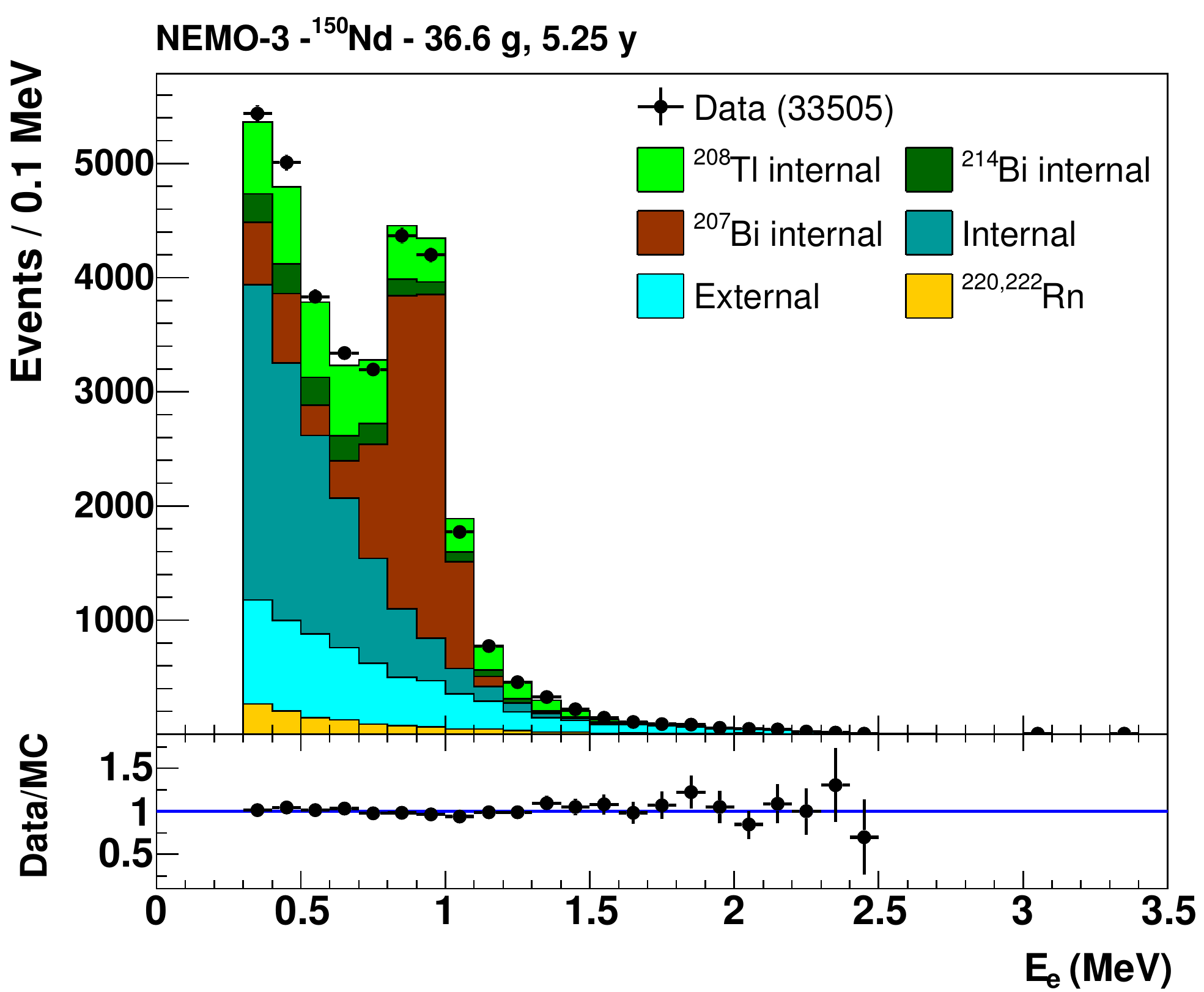}\put(200,65){\large\textbf{(a)}}\end{overpic}\label{Fig:1e1g_Ee}}\\
    \subfigure {\begin{overpic}[scale=0.4]{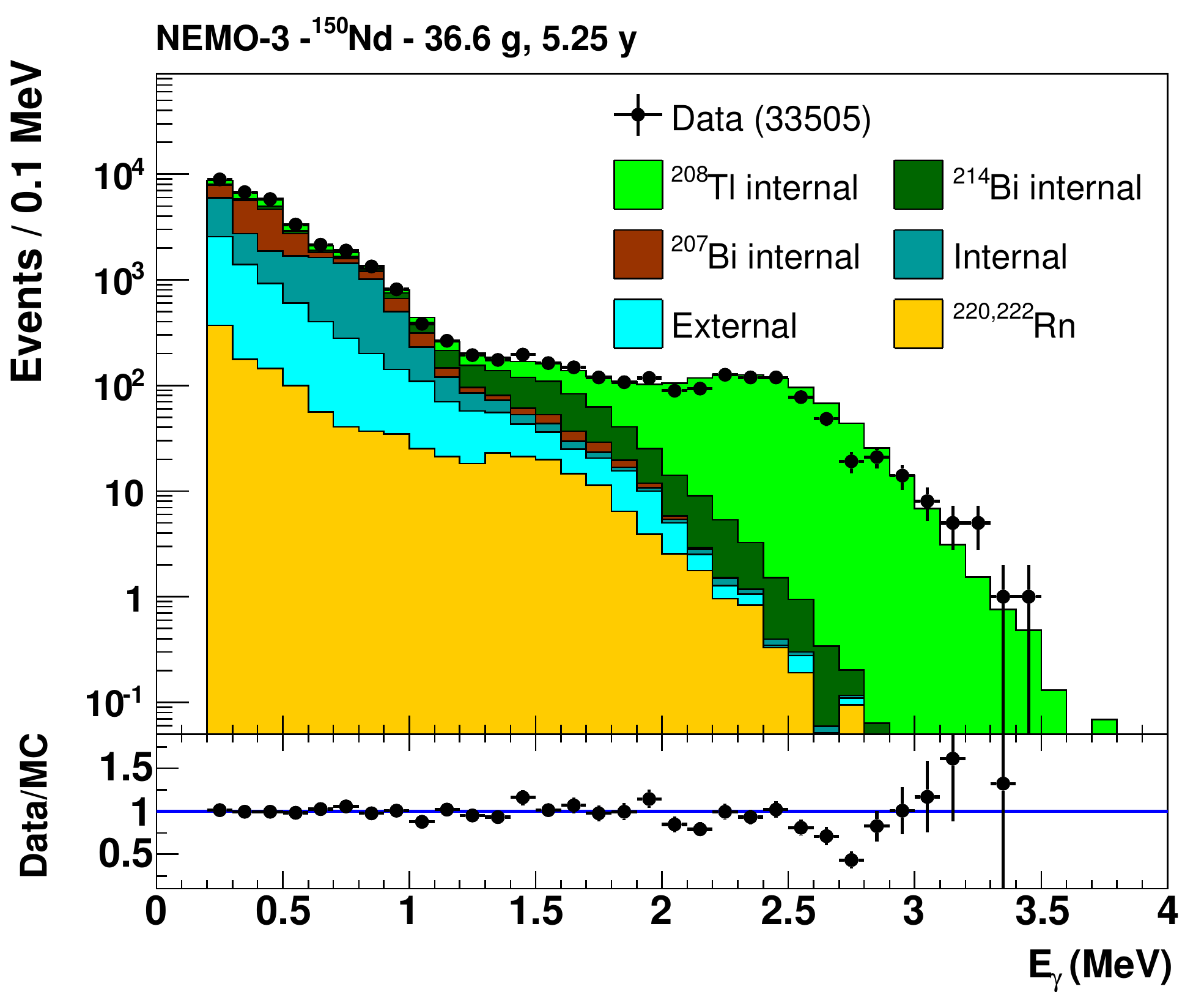}\put(200,65){\large\textbf{(b)}}\end{overpic}\label{Fig:1e1g_Eg_log}}
    \caption{\label{Fig:1eNg_distributions} Distributions of the~\subref{Fig:1e1g_Ee} electron energy ($E_e$) 
    and~\subref{Fig:1e1g_Eg_log}  $\gamma$ energy ($E_{\gamma}$) in the 1e1$\gamma$ channel for data and the background model. }
\end{figure}

Events in both the 1e1$\gamma$ and 1e2$\gamma$ channels must meet the same selection criteria as the 1e channel, with the additional requirement of one or two reconstructed $\gamma$ rays with  \unit[$E_\gamma>0.2$]{MeV}.  The time differences between the electron and each $\gamma$ hit are required to be consistent with an internal decay, such that $P_{\mathrm{int}} \geq 4$\% and $P_{\mathrm{ext}} \leq 1$\%. The $E_e$ and $E_\gamma$ distributions for events meeting the 1e1$\gamma$-channel selection criteria are shown in Fig.~\ref{Fig:1eNg_distributions}. 

\begin{figure}[htbp]
   {\includegraphics[scale=0.4]  {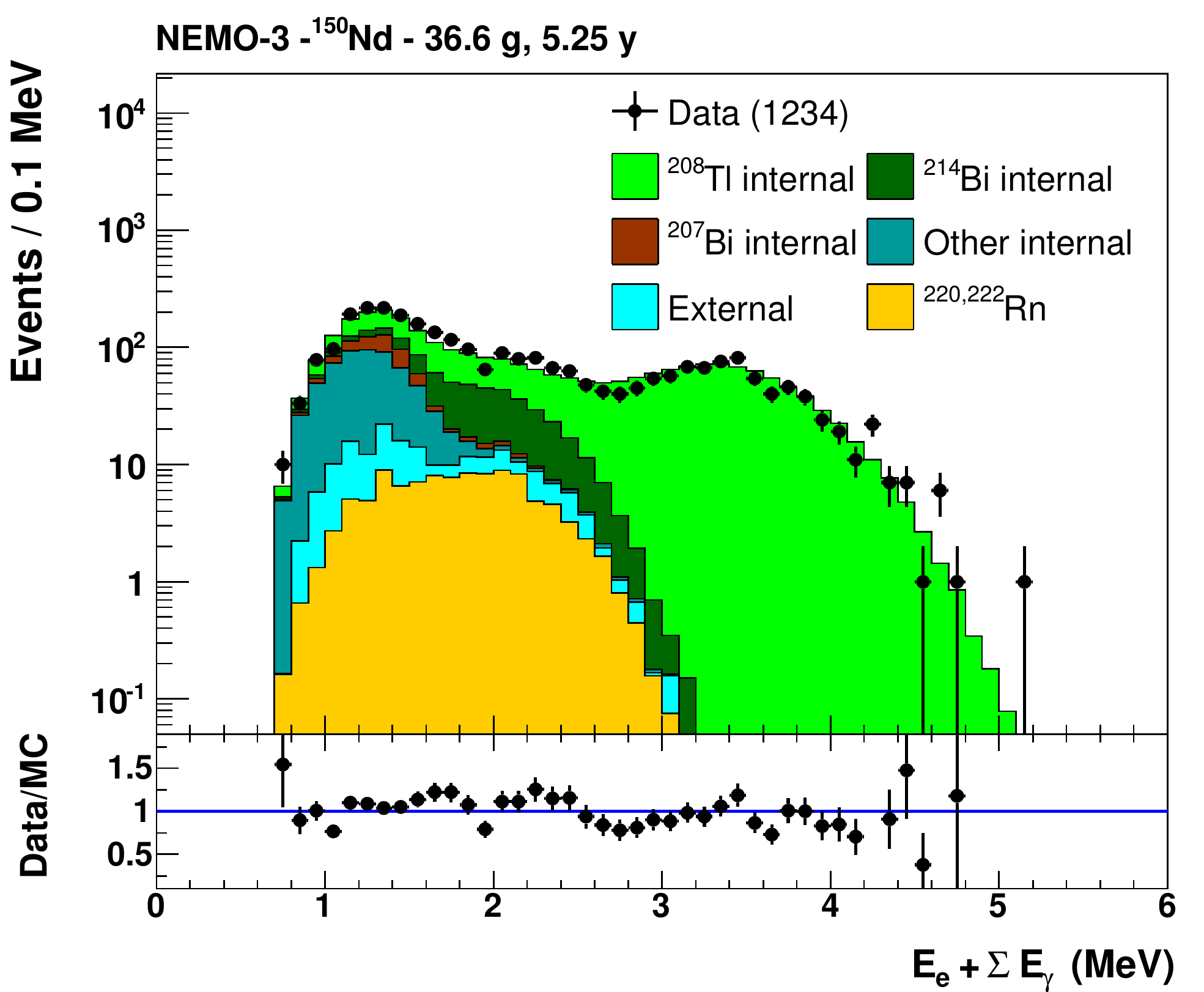}}%
    \caption{\label{Fig:1e2g_TotalE} Distribution of the total energy in the 1e2$\gamma$ channel for data and
    the background model. }
   \end{figure}
   
The electron energy spectrum in Fig.~\ref{Fig:1e1g_Ee} provides good discrimination between $^{207}$Bi and the other background decays in this channel. The conversion-electron peak from $^{207}$Bi decays is well described by the simulation. Although this background is not critical for the \zeronu search, it is an important background in the measurement of the \twonu decay half-life.  The systematic uncertainty on the normalization of $^{207}$Bi in the two-electron channel is estimated using data from special $^{207}$Bi calibration sources. The activities of these sources are measured using both the 1e(N)$\gamma$ and 2e(N)$\gamma$ channels, where $N$ can be any number between 0 and 10.  In this way, any differences between activities measured in the two channels can be attributed to the electron reconstruction efficiency, independent of the $\gamma$ reconstruction. The event selection criteria are similar to those used to select source foil events. The largest difference in activity observed between the two decay channels is $5.6\%$, which is assigned as the systematic uncertainty on the normalization of $^{207}$Bi. Varying the normalization of $^{207}$Bi by this amount has a $\pm 0.24$\% effect on the \twonu decay half-life.


The $E_\gamma$ spectrum shown in Fig.~\ref{Fig:1e1g_Eg_log} provides discrimination between $^{208}$Tl and most of the other backgrounds of interest in the 1e1$\gamma$ channel. The total energy (E$_{e}+\Sigma$E$_{\gamma}$) from the 1e2$\gamma$ channel, shown in Fig.~\ref{Fig:1e2g_TotalE}, is also used to constrain the activity of $^{208}$Tl. We observe good agreement between data and the simulation in both distributions, particularly in the high energy tails of Figs.~\ref{Fig:1e1g_Eg_log} and~\ref{Fig:1e2g_TotalE} which are dominated by $^{208}$Tl background. As equilibrium is assumed between $^{208}$Tl and $^{228}$Ac/$^{212}$Bi, which contribute events at lower $E_\gamma$, the normalization of the $^{208}$Tl background is correlated at the level of about $-10\%$ with the background from $^{214}$Bi and the europium isotopes as determined from the likelihood fit. The small deviations observed in the $E_{\gamma}$ spectrum are due to mis-modelling of high energy $\gamma$-ray interactions in the scintillator blocks. The small discrepancies are accounted for by the total systematic uncertainty of the background model.

The systematic uncertainty on the normalization of $^{208}$Tl in the two-electron channel is estimated using $^{232}$U calibration sources. The activities of the uranium sources are measured in the 1e1$\gamma$, 1e2$\gamma$ and two-electron channels using the same event selection criteria as presented in this article to select $^{208}$Tl decay events from the decay chain of $^{232}$U. The variance of the activities measured among each channel divided by their mean is $\pm 7$\%. This is taken as an estimate of the systematic uncertainty on the normalization of $^{208}$Tl.  It has a negligible effect ($< 0.1$\%) on the \twonu decay half-life measurement. However, it is important to take into account in the search for \zeronu decay of $^{150}$Nd due to the high $Q_{\beta}$ value for $^{208}$Tl of \unit[$4.999$]{MeV}~\cite{Tl208_scheme}.

The 1e1$\gamma$ channel fit prefers a higher activity for the $^{214}$Bi components of the background model compared to the 1e1$\alpha$ channel. The difference is attributed to uncertainties in the reconstruction efficiency of $\gamma$ rays and $\alpha$ particles. Mis-modelling of energy loss for $\alpha$ particles emitted from the central part of the $^{150}$Nd source foil and
inhomogeneities in the $^{214}$Bi density within the foil can result in a large variation on the $\alpha$ reconstruction efficiency.  The systematic uncertainty on the normalization of $^{214}$Bi events  is estimated from the differences between the activities measured in the 1e1$\alpha$ and 1e1$\gamma$ channels when they are fitted separately. This yields an uncertainty of $\pm 23$\% for the internal $^{214}$Bi component and $\pm 10$\% for the components in the tracker wires and foil surface.  The effect of the internal $^{214}$Bi uncertainty on the \twonu decay half-life measurement is found to be $\pm 0.3$\%, while the effect from the tracker radon is negligible ($< 0.1$\%).

The activity of $^{152}$Eu is constrained to the activity measured by HPGe detectors~\cite{TDR}, which have better sensitivity to this radionuclide. Comparing the $^{152}$Eu activity with and without this constraint leads to a difference of $14\%$, which is taken as the systematic uncertainty on its normalization. Since the decay schemes of $^{154}$Eu and $^{152}$Eu are similar, the same systematic uncertainty is applied to $^{154}$Eu and treated as completely correlated with $^{152}$Eu. The combined uncertainty on europium backgrounds has a negligible effect on the \twonu decay half-life measurement and the search for \zeronu decays.

\section{Measurement of the $2\nu\beta\beta$ decay half-life}
 \label{Sec:2vBB_measurement}

The two-electron (2e) channel provides the best sensitivity for the measurement of the $^{150}$Nd $\beta\beta$-decay rate. Candidate $\beta\beta$-decay events must have exactly two electrons that meet the selection criteria presented in Sec.~\ref{Sec:Reconstruction}. In addition, no $\alpha$ candidates are allowed within \unit[15]{cm} of the electron vertices to reduce the background from $^{214}$Bi decays.  To further improve the rejection of $^{214}$Bi background, no more than one prompt Geiger hit that is unassociated with the electron tracks is allowed within \unit[15]{cm} of the track vertices if the tracks are on opposite sides of the foil. If they are on the same side of the foil, no such hits are permitted. To improve the vertex and track reconstruction for events with electrons on the same side of the foil, each track must have a hit in the first Geiger layer closest to the source foil. This implies that the tracks are isolated from each other, since tracks can not share Geiger hits in the reconstruction. If the tracks are on the opposite sides of the source foil, they must have a hit in either the first or second Geiger layer. The timing and trajectories of the electrons are required to be consistent with the internal TOF hypothesis, i.e.,$P_{\mathrm{int}} \geq 1\%$ and $P_{\mathrm{ext}} \leq ~1\%$. The requirement $P_{\mathrm{int}} \geq 1\%$ is less stringent than for the 1e(N)$\gamma$ channel since there are two charged particles in the final state with trajectories
in the tracking detector. 

\begin{figure}[!t]
   {\includegraphics[scale=0.43]{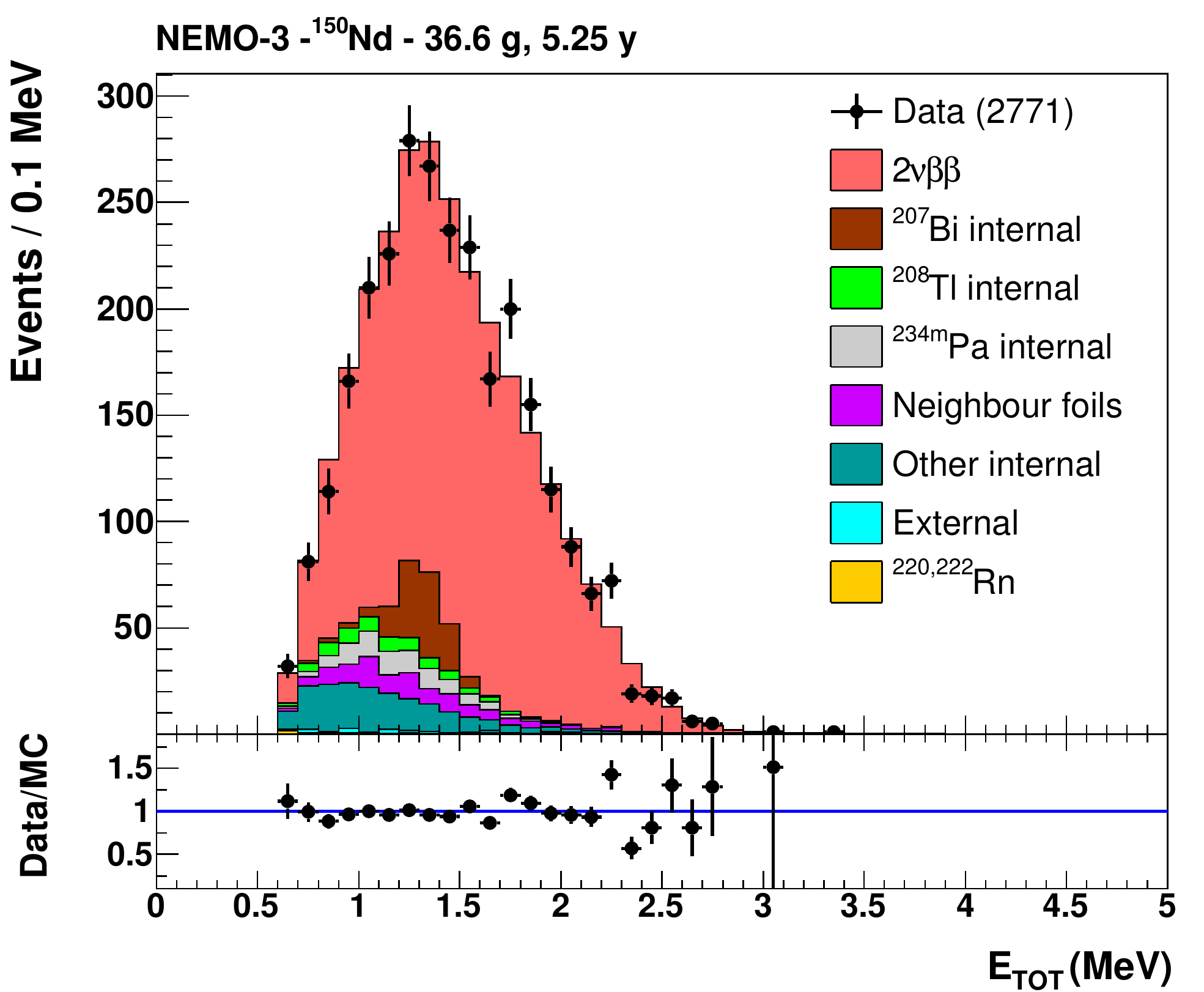}}%
    \caption{\label{Fig:2e_TotalE} Distribution of the total energy of the two electrons ($E_{\mathrm{tot}}$) after the \twonu selection. The data are compared to simulation, where the normalization of simulated events reflects the fitted activity for each component using the background model, as given in Table~\ref{Tab:2e_expected_evts}. }
\end{figure}

A total of 2771 data events from the full exposure pass the two-electron channel selection criteria.  The distribution of the total energy of both electrons ($E_{\mathrm{tot}}$) for data and MC simulation is shown in Fig.~\ref{Fig:2e_TotalE}. It provides the best discrimination between the $^{150}$Nd \twonu decay spectrum and the background decay spectra described in Sec.~\ref{Sec:Background_model}.

\begin{table*}[t]
	\centering
	\begin{tabular}{c|c|l|c}
		\hline 
		\hline 
		Channel & Observable & Processes & $\chi^{2}$/NDF\\
		\hline 
        SEC & $E_{\mathrm{tot}}$ & External backgrounds & $102 / 56\phantom{1}$ \\
        1e1$\gamma$-Ext & $E_{e}+E_{\gamma}$ & External backgrounds & $79 / 54$ \\
        1e1$\alpha$ & $\alpha$ range & $^{214}$Bi SWire, SFoil, internal backgrounds & $117 / 136$\\
        1e & $E_{e}$ & $^{210}$Bi SWire, SFoil, $^{40}$K, $^{234m}$Pa (internal) & $104 / 56\phantom{1}$ \\
        1e1$\gamma$ & $E_{\gamma}$ & $^{228}$Ac, $^{207,212,214}$Bi, $^{208}$Tl,  $^{152,154}$Eu, $^{214}$Pb (internal) & $50 / 56$ \\
        1e2$\gamma$ & $E_{e}+\Sigma E_{\gamma}$ & $^{208}$Tl (internal) & $121 / 52\phantom{1}$ \\
        2e  & $E_{\mathrm{tot}}$ & $^{150}$Nd (2$\nu\beta\beta$) & $41/ 36$ \\
		\hline
		\hline
	\end{tabular} 
    \caption{\label{Tab:Fit_components} Observables used in the binned likelihood fit for each analysis channel.  The primary isotopes of interest in each channel are also listed. The $\chi^{2}$ values and NDF~$= N_{\mathrm{bins}} -1$ are determined for each channel by comparing the data to the total number of expected events obtained from the fit. The data from Phase 1 and 2 are combined for this comparison.}
\end{table*}

We obtain the decay rates of all signal and background contributions using a binned likelihood fit to the observables from each decay channel listed in Table~\ref{Tab:Fit_components}. The only constraints in the fit are associated with the internal $^{214}$Bi contribution, which is constrained to the average activity measured in the 1e1$\alpha$ and 1e1$\gamma$ channels, and the $^{152}$Eu background, which is constrained to the rate measured by the HPGe detector. The contribution from radioactive decays in the neighboring source foils are fixed using the rates measured in dedicated analyses of these foils~\cite{Zr96_thesis,Ca48_paper,Arnold:2015wpy}.  All other background rates and the \twonu decay rate are fitted without constraints over a total of $446$ bins with $32$ parameters. We observe strong correlations  between several sources of background as discussed throughout Sec.~\ref{Sec:Background_model}. 
The \twonu decay signal normalization, however, is largely uncorrelated with most backgrounds. The strongest correlation coefficient observed is $-11\%$ between the signal and the $^{234m}$Pa background. 

We perform the likelihood minimization using only the number of counts in each bin. The impact of the systematic uncertainties on the \twonu decay half-life is determined independently. The data are compared to the sum of the MC components at their best fit normalization in all channels using a $\chi^{2}$ method, which yields a $\chi^{2} = 614$ with a number of degrees of freedom (NDF) of 414.  Table~\ref{Tab:Fit_components} contains a breakdown of the $\chi^{2}$ contributions from each channel used in the fit. The data in each channel are consistent with the expectation from the MC simulation. The level of agreement between data and MC expectation varies slightly across the channels, which is due to the large statistical variations in sample sizes as well as the systematic uncertainties on the normalization of some backgrounds, which are not included in this fit.  

The fitted activities and expected numbers of events in the two-electron channel from each background source are given in Table~\ref{Tab:2e_expected_evts}. We compare the activities of the internal backgrounds to the HPGe measurements of the source foil contamination to validate the modelling of the detector response and the background. For most isotopes, we observe good agreement between the HPGe and \nemo detector results within their total uncertainties. 

The $^{214}$Bi activity extracted from the global S+B fit is significantly larger than the upper limit obtained with the HPGe detector at the $90\%$ C.L.
The \nemo result is cross checked in multiple decay channels. The observed differences are accounted for by the $23\%$ systematic uncertainty estimated for the normalization of this background. Taking this systematic uncertainty into account, the \nemo result is within $2 \sigma$ of the upper limit on the $^{214}$Bi activity from the HPGe measurements. As it is not an important background in the search for \zeronu decay in the $^{150}$Nd foil, this discrepancy has no impact on the results of this analysis. The $^{207}$Bi activity measured by \nemo is lower than expected from the HPGe measurements due to the removal of the hot spots described in Sec.~\ref{Sec:Reconstruction}.  Without removal of these hot spots, the $^{207}$Bi activity measured is \unit[$132.8 \pm 6.8$]{mBq/kg} in the 1e1$\gamma$ channel, which is consistent with the HPGe result. Finally, the measured activity of $^{40}$K is nearly a factor of $2.5$ larger than the HPGe limit. The HPGe detector is not efficient in detecting the high energy (\unit[1.46]{MeV}) $\gamma$ ray produced in the electron capture of $^{40}$K. In addition, it is possible that some additional $^{40}$K contamination may have been introduced on the foil surface during the construction of the \nemo detector. Therefore, the \nemo result is preferred over the measurement from the HPGe detectors.

We estimate the systematic uncertainties on the numbers of events in the two-electron channel from $^{208}$Tl ($^{228}$Ac, $^{212}$Bi), $^{214}$Bi, $^{152,154}$Eu, and $^{207}$Bi decays using independent control channels and calibration sources as discussed in Sec.~\ref{Sec:Background_model}.  The systematic uncertainty on the total number of events from internal $^{40}$K and $^{234m}$Pa decays is estimated using the two-electron channel. 
As $^{40}$K and $^{234m}$Pa are both predominantly $\beta$ emitters whose decay rates are constrained by the one-electron channel, they suffer from similar systematic uncertainties in two-electron channel events. The source of the uncertainty is due to a combination of mis-modelling of M\o ller scattering, energy loss, bremsstrahlung and tracking within the simulation and reconstruction.  The systematic uncertainty for these isotopes is estimated by fitting their total contribution to the $E_{\mathrm{tot}}$
distribution of the two-electron channel,
with their relative normalizations and the normalizations of all other components fixed to their best fit values. This results in a reduction in the $^{40}$K and $^{234m}$Pa total event rate that is treated as a symmetric $\pm 10$\% systematic uncertainty on their normalization in the two-electron channel, leading to an uncertainty of $\pm 2.5$\% on the measured $T_{1/2}^{2\nu}$.   

The systematic uncertainty on the number of events from neighboring foil isotopes is estimated in a similar way. Allowing their total contribution to float with all other contributions fixed yields a reduction of the neighboring foil activity by $23\%$. Treated as symmetric, this translates into a $\pm 0.5$\% effect on the \twonu decay rate. The effect of all background normalization uncertainties contributes $[+2.8, -2.7]\%$ to the total systematic uncertainty on the $T_{1/2}^{2\nu}$ measurement.

\begin{table}[t!]
	\centering
	\begin{tabular}{c|cc|cc}
	\hline
	\hline
	\multirow{2}{*}{Contribution}&\multicolumn{2}{c|}{ \text{\unit[Activity]{(mBq/kg) } }} &\multirow{2}{*}{$N_{2e}$}\\ 
			            & \multicolumn{1}{c}{HPGe} & \multicolumn{1}{c|}{\nemo} &   \\
		\hline 
		$^{207}$Bi & $130 \pm 5$ & $99.6 \pm 2.6$  & $130.4 \pm 3.4  \pm 7.3$ \\
		Neighboring foils &  \multicolumn{1}{c}{} & \multicolumn{1}{c|}{ } & $96.8 \pm 22.1$  \\	
		$^{234m}$Pa & $< 66$  & $27.0 \pm 0.6$  & $80.8 \pm 1.8  \pm 7.9$  \\
		$^{208}$Tl & $10 \pm 2$  & $10.9 \pm 0.2$  & $60.3 \pm 1.2  \pm 4.2$  \\
		$^{228}$Ac & $20 \pm 7$  & $30.3 \pm 0.6$  & $58.3 \pm 1.2  \pm 4.1$ \\		
		$^{40}$K & $< 70$ & $179.0 \pm 1.4 \phantom{1}$  & $50.9 \pm 0.4  \pm 5.0$ \\
		$^{212}$Bi & $20 \pm 7$  & $30.3 \pm 0.6$  & $31.4 \pm 0.6  \pm 2.2$ \\
		$^{214}$Bi & $< 3 $ & $\phantom{1}5.1 \pm 0.2$  &  $26.9 \pm 1.3  \pm 6.1$ \\
		External &  \multicolumn{1}{c}{ } & \multicolumn{1}{c|}{ } & $15.1 \pm 0.4 ~^{+~4.6}_{-~3.5}$  \\		
		$^{220,222}$Rn & \multicolumn{1}{c}{ } & \multicolumn{1}{c|}{ } & $\phantom{1}8.0 \pm 1.0  \pm 0.8$ \\
		$^{154}$Eu & \multicolumn{1}{c}{ } & $19.1 \pm 4.3$  & $\phantom{1}4.4 \pm 1.0  \pm 0.6$ \\
		$^{152}$Eu & $40 \pm 5$  &  $52.5 \pm 3.6$  & $\phantom{1}2.2 \pm 0.2  \pm 0.3$ \\
		$^{214}$Pb & $< 3$  &  $\phantom{1}5.1 \pm 0.2$  & $\phantom{1}0.6 \pm 0.1  \pm 0.1$\\			\hline
		Total Background &  \multicolumn{1}{c}{ } & \multicolumn{1}{c|}{ } & $566.2 \pm 4.9 ~^{+~29.6}_{-~29.5}$\\	
		Data &  \multicolumn{1}{c}{ } & \multicolumn{1}{c|}{ } & 2771  \\	
		\hline 
		\hline
	    \end{tabular} 
	  \caption{\label{Tab:2e_expected_evts}  The activities measured with the \nemo data using the \twonu selection criteria are compared to HPGe measurements of the $^{150}$Nd source foil~\cite{TDR}, for which only the $\pm1$ standard deviation statistical uncertainties are shown. There is an additional systematic uncertainty of $10\%$ on the HPGe measurements. Secular equilibrium between $^{214}$Bi and $^{214}$Pb is assumed. The same assumption is made for $^{208}$Tl, $^{228}$Ac, and $^{212}$Bi, where the branching ratio of $35.94\%$ is taken into account. Events from $^{207}$Bi and $^{152,154}$Eu are weighted by their respective half-lives as a function of the event time and are therefore reported as of February 1, 2003. The corresponding numbers of expected events in the two-electron channel $(N_{2e})$ from each background are also provided with statistical and systematic uncertainties.} 
\end{table}

Several additional sources of systematic uncertainty are investigated. The largest source of uncertainty is associated with the absolute normalization of $\varepsilon_{2e}$ (see Eq.~\ref{Eq:half_life}). This uncertainty is estimated through the comparison of $^{207}$Bi calibration source activities measured with the NEMO-3 detector and an HPGe detector. Using the 2e(N)$\gamma$ channel to measure these sources yields activities that are in agreement with the HPGe measurements to within $\pm 5.55 \%$, which is taken as the uncertainty on $\varepsilon_{2e}$. 

The effects of incorrect simulation of energy loss and bremsstrahlung on $\varepsilon_{2e}$ are also investigated. We generate MC data sets with these various parameters altered within their expected uncertainty. The resulting effects on $T_{1/2}^{2\nu}$ are found to be on the order of $1\%$ or less for each of the individual sources of uncertainty (See Table~\ref{Tab:2vBB_systematics}). 

The $^{150}$Nd source foil is the thinnest composite foil produced for the \nemo experiment, and the composite powder itself was of very good quality with fine granularity.  Nevertheless, there remains some level of uncertainty about the homogeneity of the source foil density.  This effect is assessed by varying the thickness of the source foil within the MC simulation, and through numerical calculations given the particulate size in the powder. The
corresponding uncertainties on $T_{1/2}^{2\nu}$ are found to be negligible ($<1$\%). The uncertainty on the foil position and rotation relative to the simulated foil radius also have a negligible effect.

The $\pm 0.5\%$ uncertainty on the enrichment factor~\cite{TDR} translates into the same uncertainty on $T_{1/2}^{2\nu}$ and the $1\%$ uncertainty on the energy calibration leads to an uncertainty of $[+1.48, -1.50]\%$ on $T_{1/2}^{2\nu}$. The systematic uncertainties summarized in Table~\ref{Tab:2vBB_systematics} result in a total systematic uncertainty of $[+6.59, -6.45]\%$.

The fitted number of $\beta\beta$-decay events from $^{150}$Nd in the two-electron channel is $N=2214.0\pm52.3$~(stat), where the uncertainty is propagated from the global likelihood fit.  This number of events is translated into a half-life using
\begin{equation}
\label{Eq:half_life}
T_{1/2}^{2\nu} =  \frac{ \ln 2 \cdot t \cdot \varepsilon_{2e} \cdot N_{\mathrm{nuclei}}}{N} \,
\end{equation}
where $t$ is the total live time of the experiment, $\varepsilon_{2e}$ is the efficiency for selecting two-electron events, and $N_{\mathrm{nuclei}}$ is the number of $^{150}$Nd atoms. Given the \twonu decay selection efficiency of $\varepsilon_{2e} = 3.87$\% and an exposure of \unit[0.19]{kg$\cdot$yr}, the \twonu decay half-life for $^{150}$Nd is measured to be
\begin{equation}\label{Eq:2vBB_halflife}
T^{2\nu}_{1/2} = [9.34 \pm 0.22~\mathrm{(stat)}~^{+0.62}_{-0.60}~\mathrm{(syst)}] \times 10^{18}\mathrm{y},
\end{equation} 
with a signal-to-background ratio of $3.9$. This result is approximately $2.4\sigma$ lower than the half-life measured by ITEP~\cite{MoscowTPC} and $2.6\sigma$ higher than the result obtained by the Irvine group~\cite{PhysRevC.56.2451}. The half-life presented herein represents the most accurate measurement of the \twonu decay half-life to date for this isotope due to a more thorough assessment of systematic uncertainties, leading to a relatively small reduction in the systematic uncertainty of $\approx 5.6\%$ compared to Ref.~\cite{Nd150_pub}.  Although the live time of the data has approximately doubled since~\cite{Nd150_pub}, the statistical uncertainty has not improved significantly. This is due to an overall lower efficiency in accepted $\beta\beta$-decay events. This reduction in signal efficiency is caused by the increased minimum energy requirement for the electrons in addition to several more strict data quality criteria applied to the data. In particular, the use of the laser survey to reject events involving unstable PMTs has a large impact on the final selection efficiency~\cite{Arnold:2015wpy}. Although these more stringent selection criteria have reduced the signal efficiency, they also provide a more stable sample of 2e events where the systematic uncertainties are better understood. 

\begin{table}[h]
	\centering
	\begin{tabular}{c|c}
	\hline
	\hline
		Source of uncertainty & Effect on $T_{1/2}^{2\nu} (\%)$\\ 
        \hline
		Absolute normalization of $\varepsilon_{2e}$  & $\pm$5.55  \\
		Foil granularity & $\pm$0.45 \\
		Foil thickness & $\pm$0.73 \\
		Enrichment & $\pm$0.50\\
		Energy loss & $\pm$0.49 \\
		Bremsstrahlung modelling & $[+1.12, -0.50]$\\
		Energy calibration & $[+1.48, -1.50]$\\
		Total background & $[+2.83, -2.66]$\\
		\hline
		Total & $[+6.59, -6.45]$\\
		\hline
		\hline
	\end{tabular} 
    \caption{\label{Tab:2vBB_systematics} Sources of systematic uncertainty on the measurement of the $2\nu\beta\beta$ half-life.  The dominant source of uncertainty is the absolute normalization of the two-electron channel efficiency for reconstructing two-electron decays.}
\end{table}

\section{Search for $0\nu\beta\beta$ decays}
\label{Sec:0vBB_limits}

We perform a search for three types of \zeronu decay modes for which the observable particle content of the final state is the same as in \twonu decay. The first involves the exchange of a light Majorana neutrino and is referred to as the mass mechanism. The rate (see Eq.~\ref{Eq:halflife_zeronu}) of this decay is related to the effective Majorana neutrino mass, $\langle m_{\nu} \rangle$, as described in Sec.~\ref{Sec:Introduction}. 

The second category involves decays mediated by right-handed currents (RHCs). Two distinct RHC mechanisms are investigated. The first RHC decay mode involves decays with a pure $W_{R}$ boson exchange at one vertex, and thus the rate depends on the coupling $\langle \lambda \rangle$ between right-handed (RH) quarks and RH leptons ($\xi = \langle \lambda \rangle$ in Eq.~\ref{Eq:halflife_zeronu}). The other RHC mechanism involves the exchange of a $W$ boson that is a mixture of $W_{R}$ and $W_{L}$ states. Its rate is given by the coupling $\langle \eta \rangle$ between left-handed (LH) quarks and RH leptons ($\xi  = \langle \eta \rangle$ in Eq.~\ref{Eq:halflife_zeronu}). 

A \zeronu decay may also proceed via the emission of majorons ($\chi^{0}$),
which are weakly interacting, light or massless bosons present in many grand unified theories. They can couple weakly to the neutrino and participate in \zeronu decay~\cite{Bamert:1994hb,Carone:1993jv,Mohapatra:2000px}. Majorons are not detected with the \nemo detector. Therefore, the total electron energy in \zeronu decays involving the emission of a majoron will form a continuous spectrum, similar to that of \twonu decay. The shape of the total energy spectrum depends on the available phase space through the relation $G^{0\nu\chi^{0}_{n}} \propto (Q_{\beta\beta} - E_{\mathrm{tot}})^{n}$, where $n$ is a spectral index. We search for \zeronu decays involving the emission of a single majoron with a spectral index of $n = 1$.

The two-electron channel is used to search for the \zeronu decay of $^{150}$Nd to the ground state of $^{150}$Sm. A looser event selection, allowing both positively and negatively charged particle tracks in the 2e channel, is adopted to improve the signal sensitivity for this decay, since tracks associated with high energy electrons have a larger radius of curvature and are therefore more susceptible to charge mis-identification. 
The global S+B likelihood fit to data meeting this looser selection criteria yields background activities that do not deviate significantly from the values in Table~\ref{Tab:2e_expected_evts}. The measured \twonu half-life is also consistent with the results of Eq.~\ref{Eq:2vBB_halflife}. This new selection increases the ratio of expected signal over background in the region of interest, thereby improving the sensitivity to \zeronu decay.

\begin{figure}
   {\includegraphics[scale=0.4]{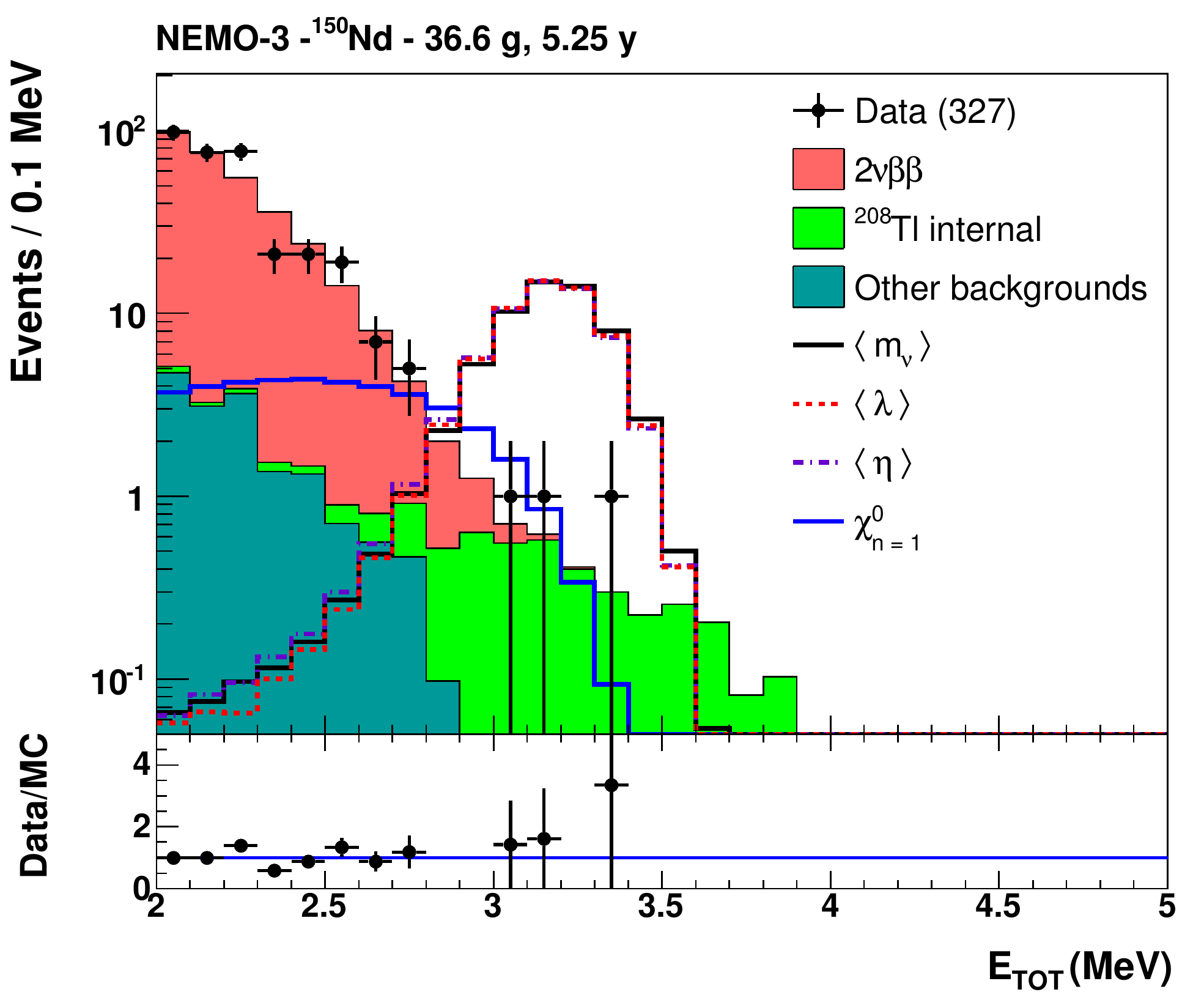}}
   \caption{\label{Fig:2e_TotalE_zoom} Distribution of the total energy in the two-electron channel for
   $E_{\mathrm{tot}}$. The signal shapes for the various \zeronu decay mechanisms investigated are shown with the same arbitrary normalizations to highlight the region of interest near the $Q_{\beta\beta}$ value for most mechanisms. The main backgrounds at the endpoint of the $\beta\beta$ spectrum come from the decay of $^{208}$Tl and \twonu decays from $^{150}$Nd.}
\end{figure}

The kinematics of each of the two electrons produced in \zeronu decays differ among the investigated mechanisms. In the case of the mass mechanism and for decays involving RHCs, the total energy of the two electrons is equal to the $Q_{\beta\beta}$ value of $^{150}$Nd, as there are no other particles involved in the decay. Figure~\ref{Fig:2e_TotalE_zoom} shows the expected signal distributions in the high energy tail of the $E_{\mathrm{tot}}$ distribution of the two electrons. The signals are shown with an arbitrary normalization compared to the expected backgrounds. There are $6$ expected background events with \unit[$E_{\mathrm{tot}}>2.8$]{MeV}, 
where $\approx 37\%$ are from \twonu decays of $^{150}$Nd, $60\%$ from internal $^{208}$Tl decays, and $2\%$ from internal $^{214}$Bi decays. Other internal background decays and radon-progeny decays in the tracker make up the remaining $1\%$.

\begin{figure*}
    \subfigure{\begin{overpic}[scale=0.4]{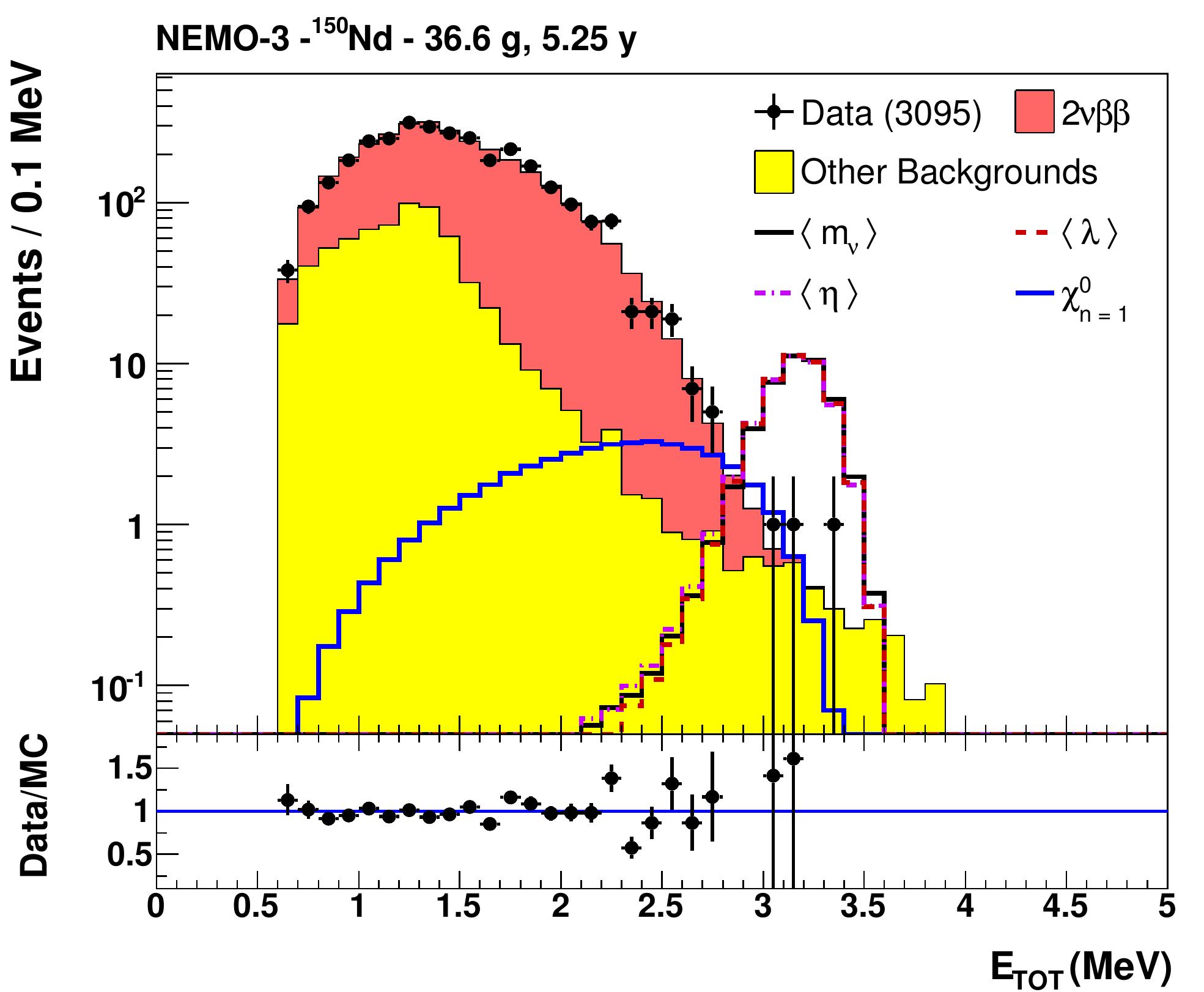}\put(200,60){\large\textbf{(a)}}\end{overpic}~\label{Fig:2e_TotalE_simple}}%
    \subfigure{\begin{overpic}[scale=0.4]{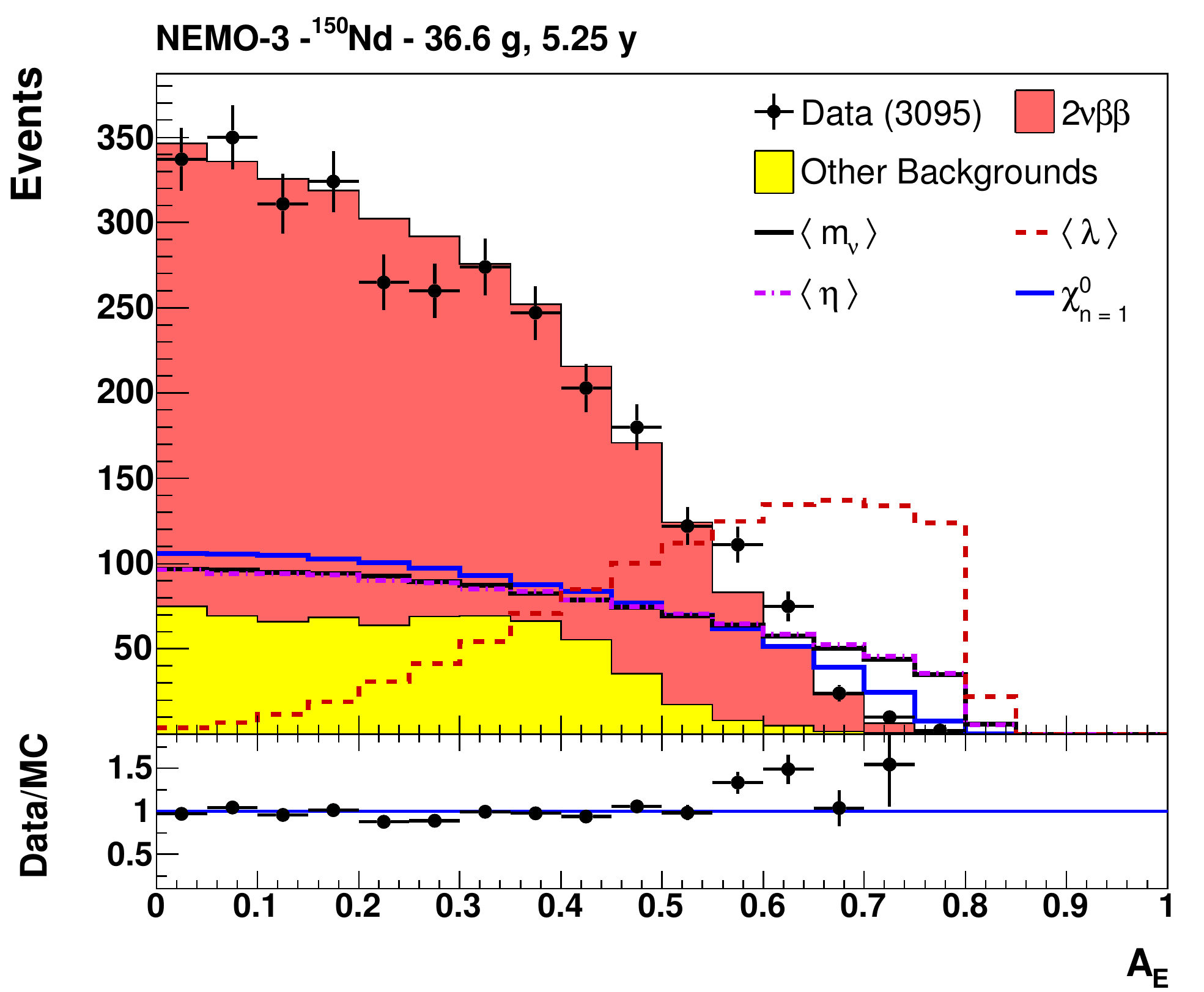}\put(200,60){\large\textbf{(b)}}\end{overpic}~\label{Fig:2e_Asymm}}\\
    \subfigure{\begin{overpic}[scale=0.4]{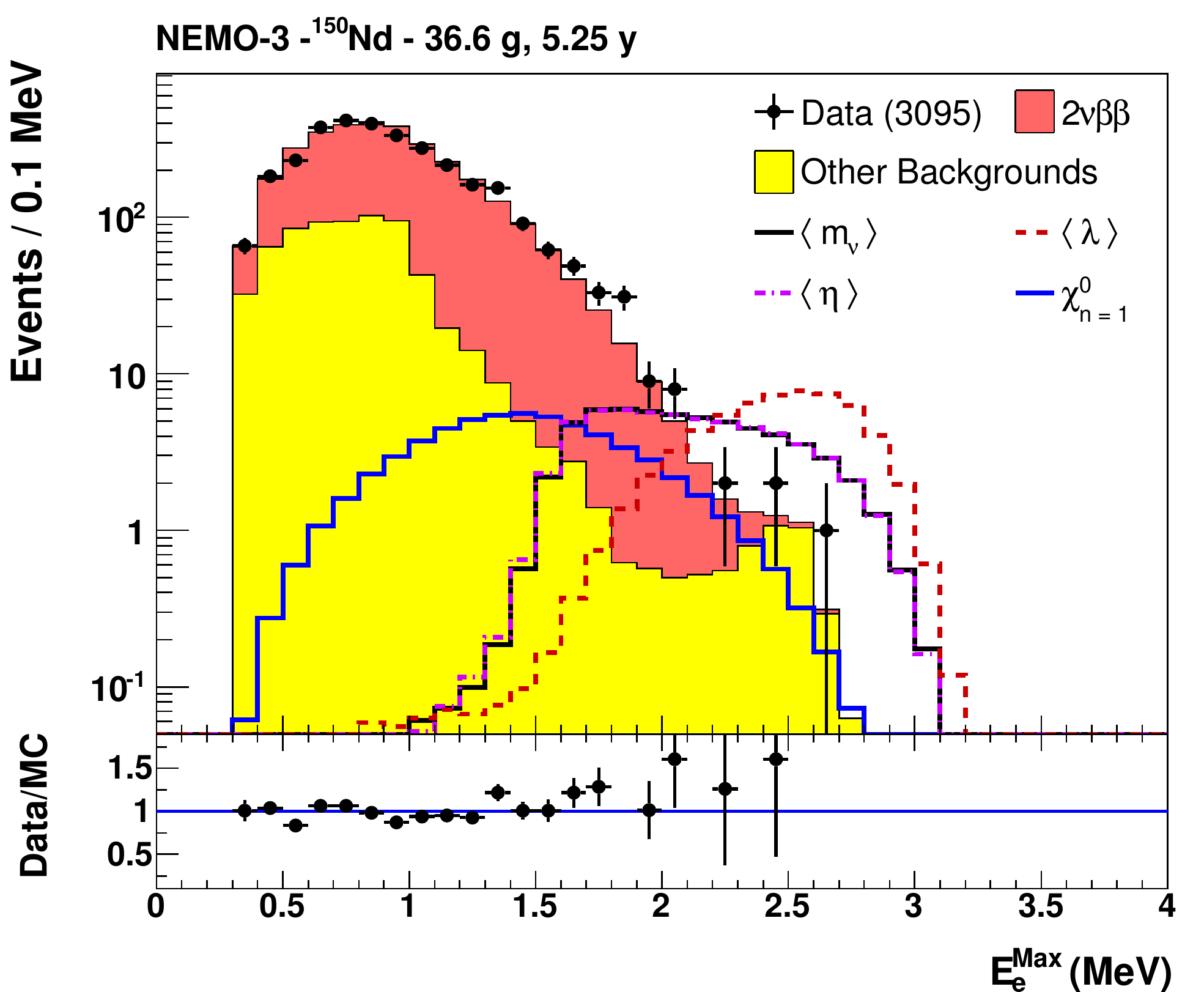}\put(200,60){\large\textbf{(c)}}\end{overpic}~\label{Fig:2e_EeH}}%
    \subfigure{\begin{overpic}[scale=0.4]{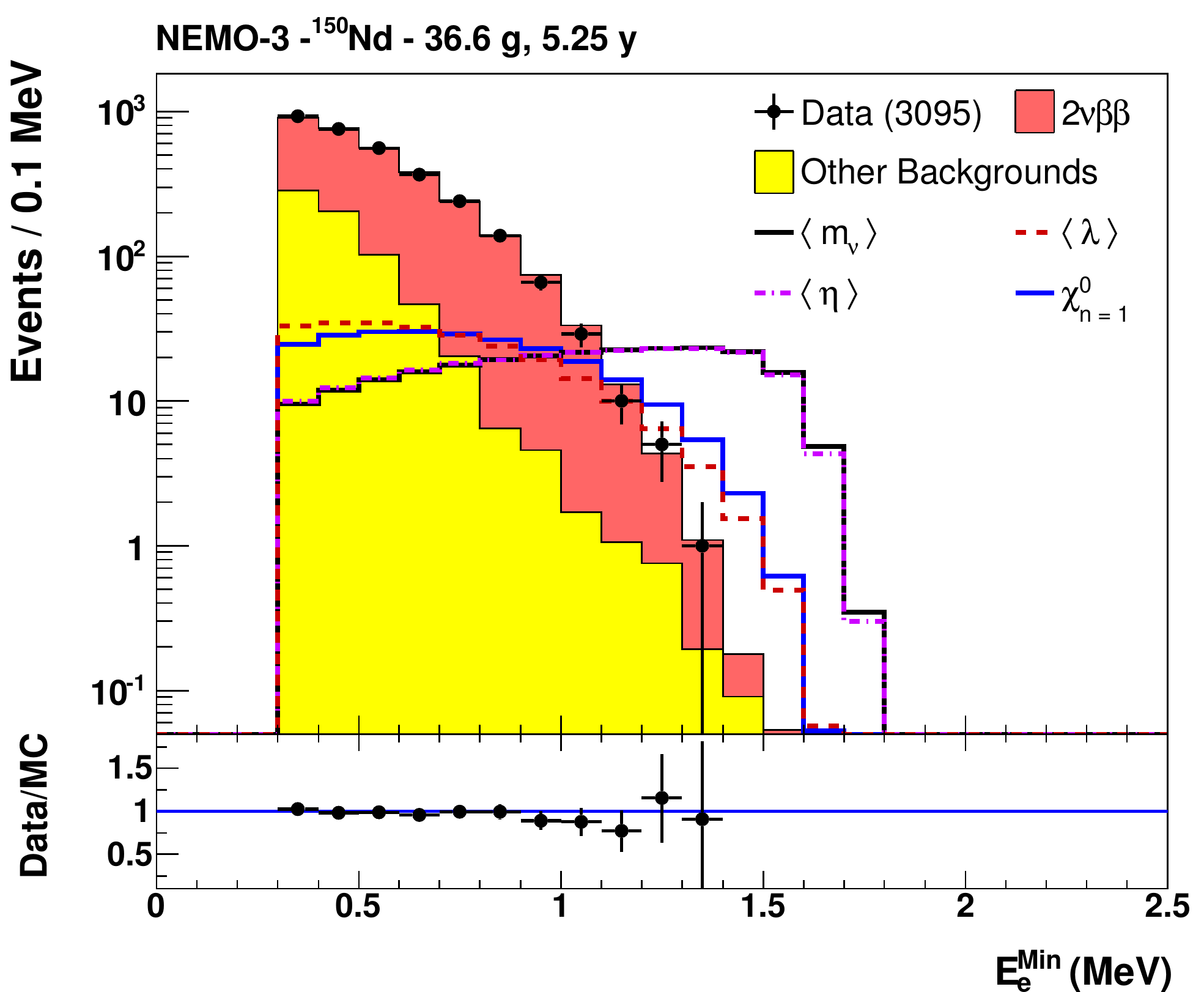}\put(200,60){\large\textbf{(d)}}\end{overpic}~\label{Fig:2e_EeL}}%
       \caption{\label{Fig:2e_channel_observables1} Distributions of \subref{Fig:2e_TotalE_simple} the total energy, \subref{Fig:2e_Asymm} the normalized energy asymmetry, and the individually reconstructed energies of the \subref{Fig:2e_EeH} higher and \subref{Fig:2e_EeL} lower energy electron from the two-electron channel. The data are compared to the total expected background, where the normalization of simulated events reflects the fitted number of events from the global S+B fit without the negative track curvature requirement. The shapes of \zeronu decay signals assuming several possible underlying mechanisms ($\langle m_{\nu} \rangle$, $\langle \lambda \rangle$, $\langle \eta \rangle$ and $\chi^{0}_{n = 1}$) are also shown with an arbitrary normalization in each figure to demonstrate the kinematic differences among the various mechanisms, as well as the potential discriminating power between each signal and the backgrounds.}
\end{figure*}

\begin{figure*}
    \subfigure{\begin{overpic}[scale=0.4]{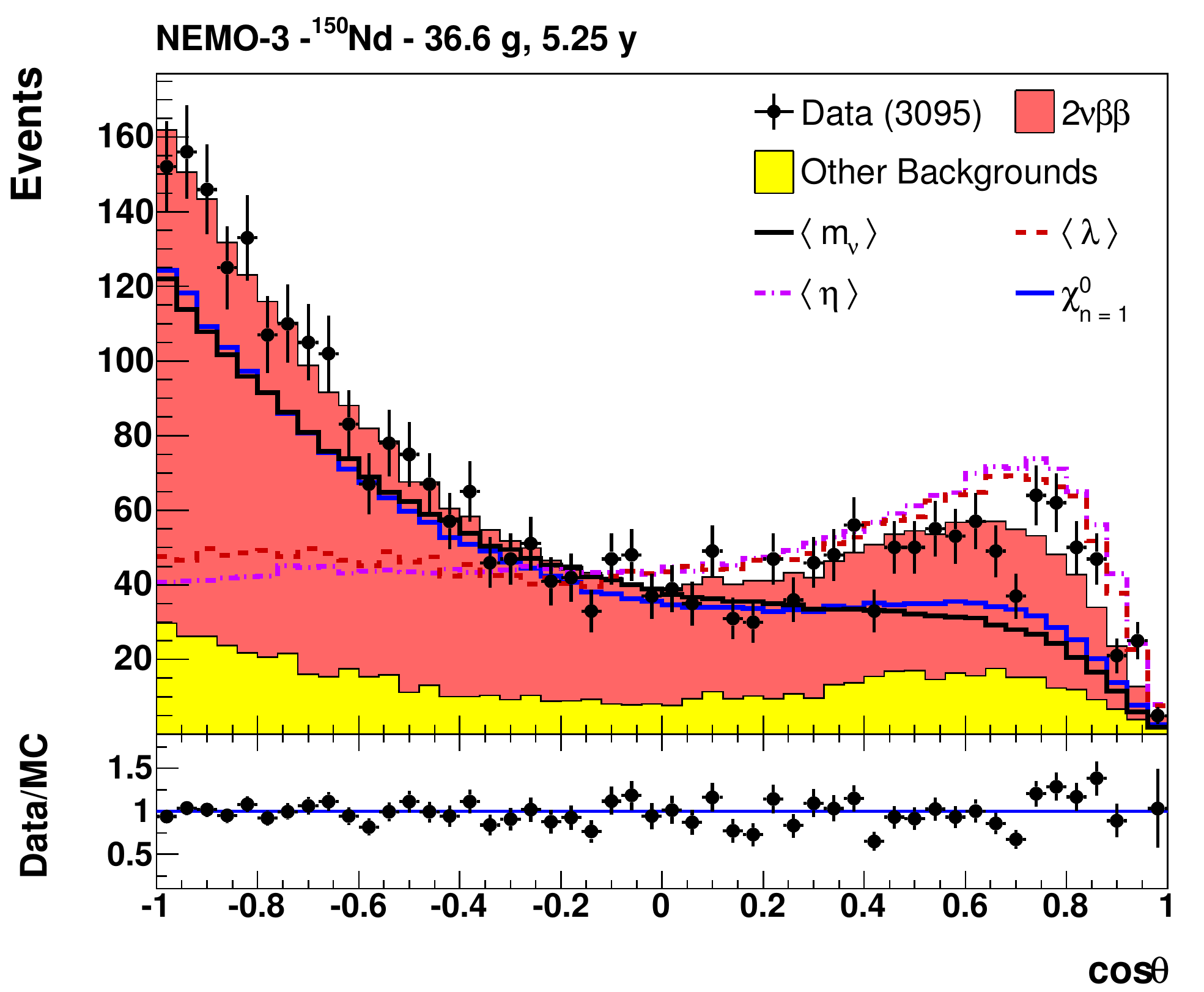}\put(40,160){\large\textbf{(a)}}\end{overpic}~\label{Fig:2e_Cosee}}%
    \subfigure{\begin{overpic}[scale=0.4]{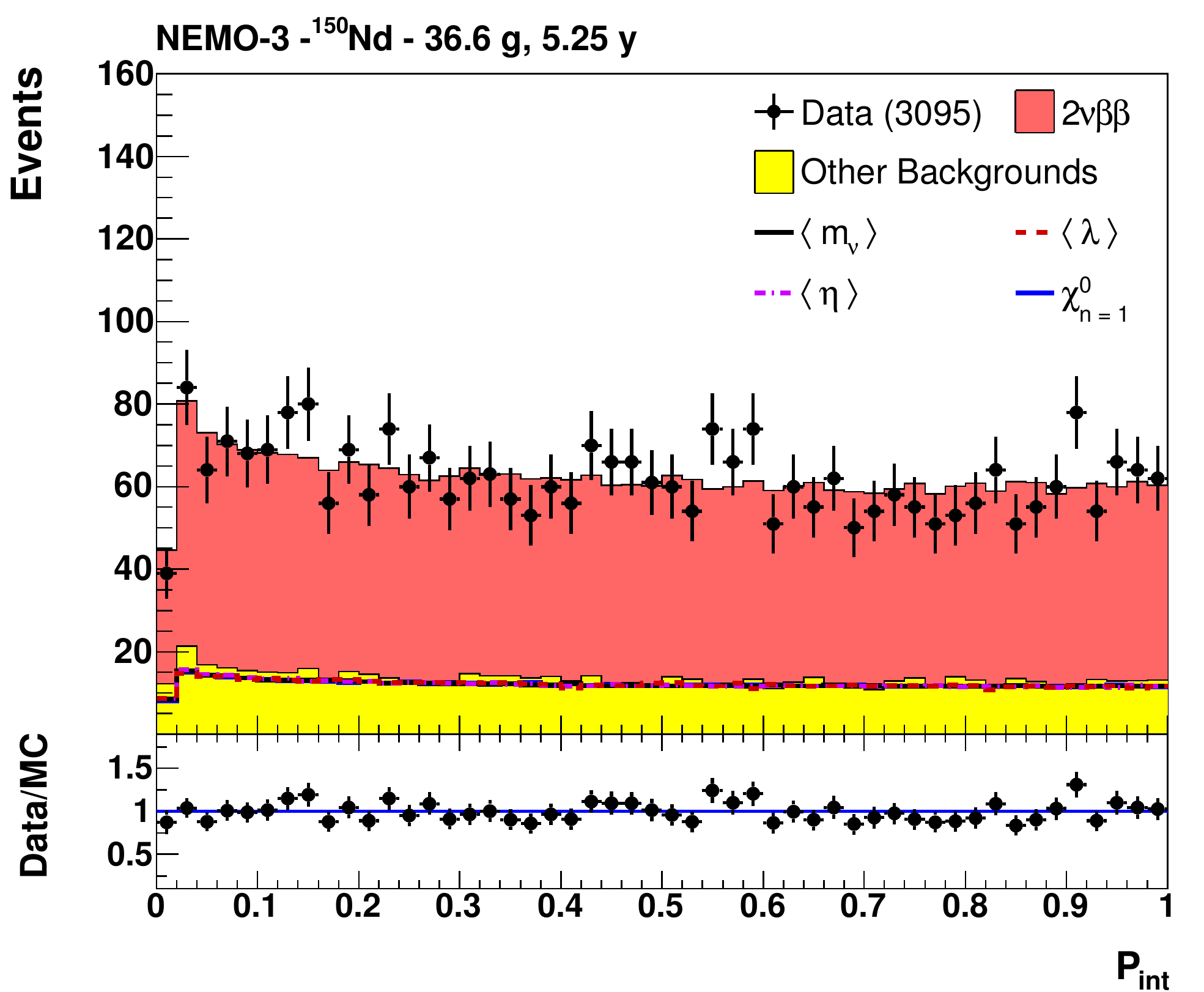}\put(40,160){\large\textbf{(b)}}\end{overpic}~\label{Fig:2e_Pint}} \\ 
    \subfigure{\begin{overpic}[scale=0.4]{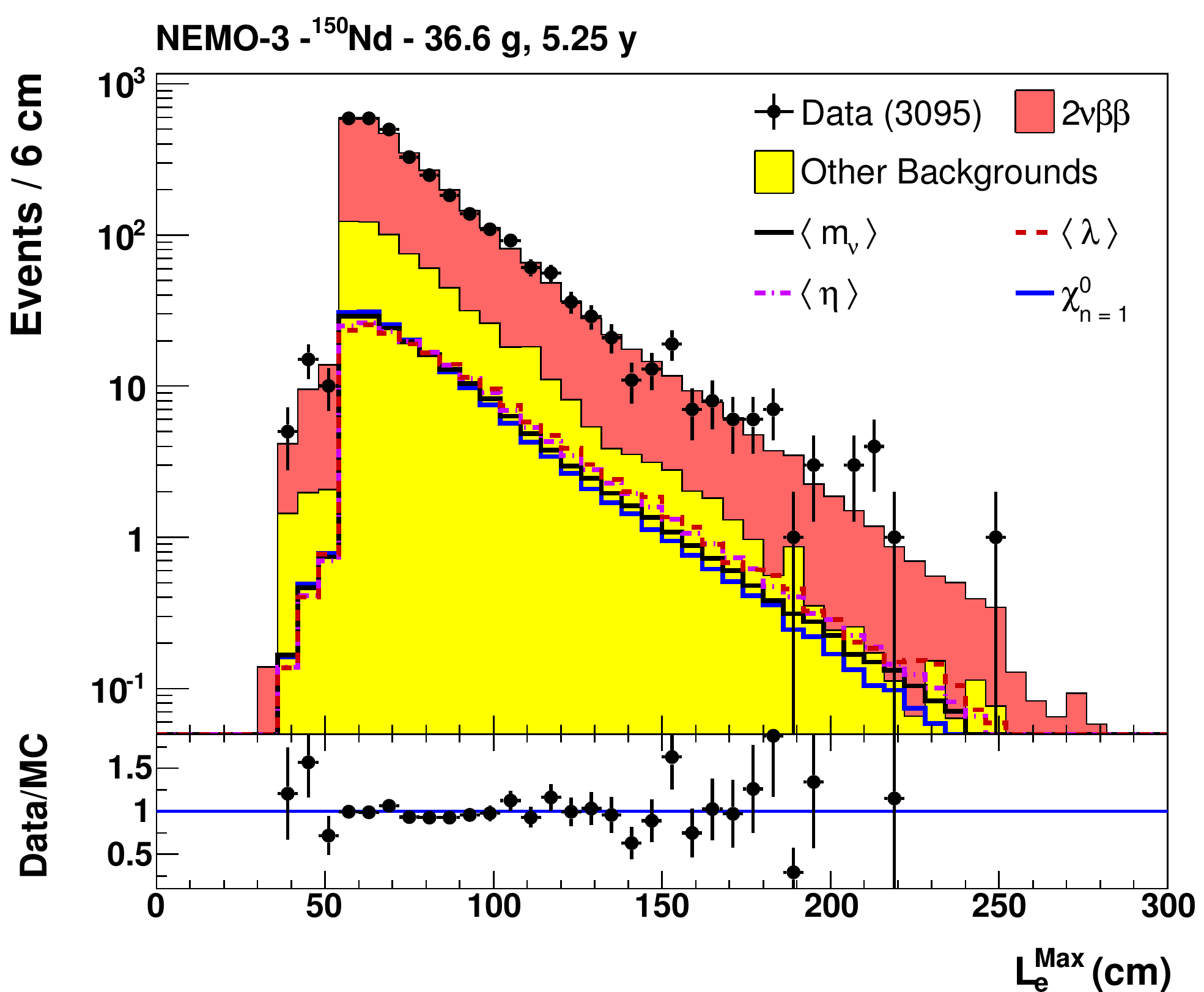}\put(40,160){\large\textbf{(c)}}\end{overpic}~\label{Fig:2e_LeMax}}%
    \subfigure{\begin{overpic}[scale=0.4]{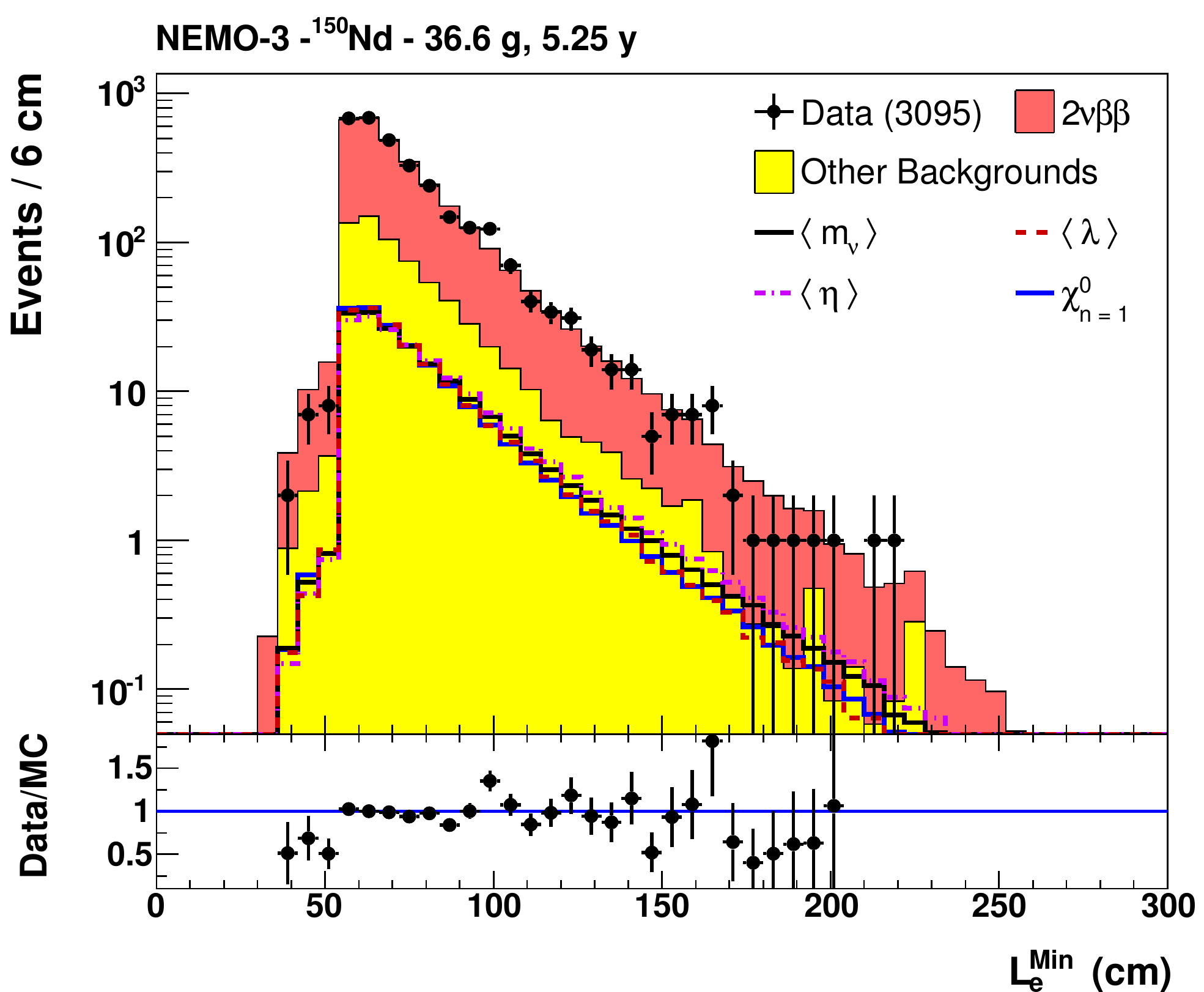}\put(40,160){\large\textbf{(d)}}\end{overpic}~\label{Fig:2e_LeMin}}
    \subfigure{\begin{overpic}[scale=0.4]{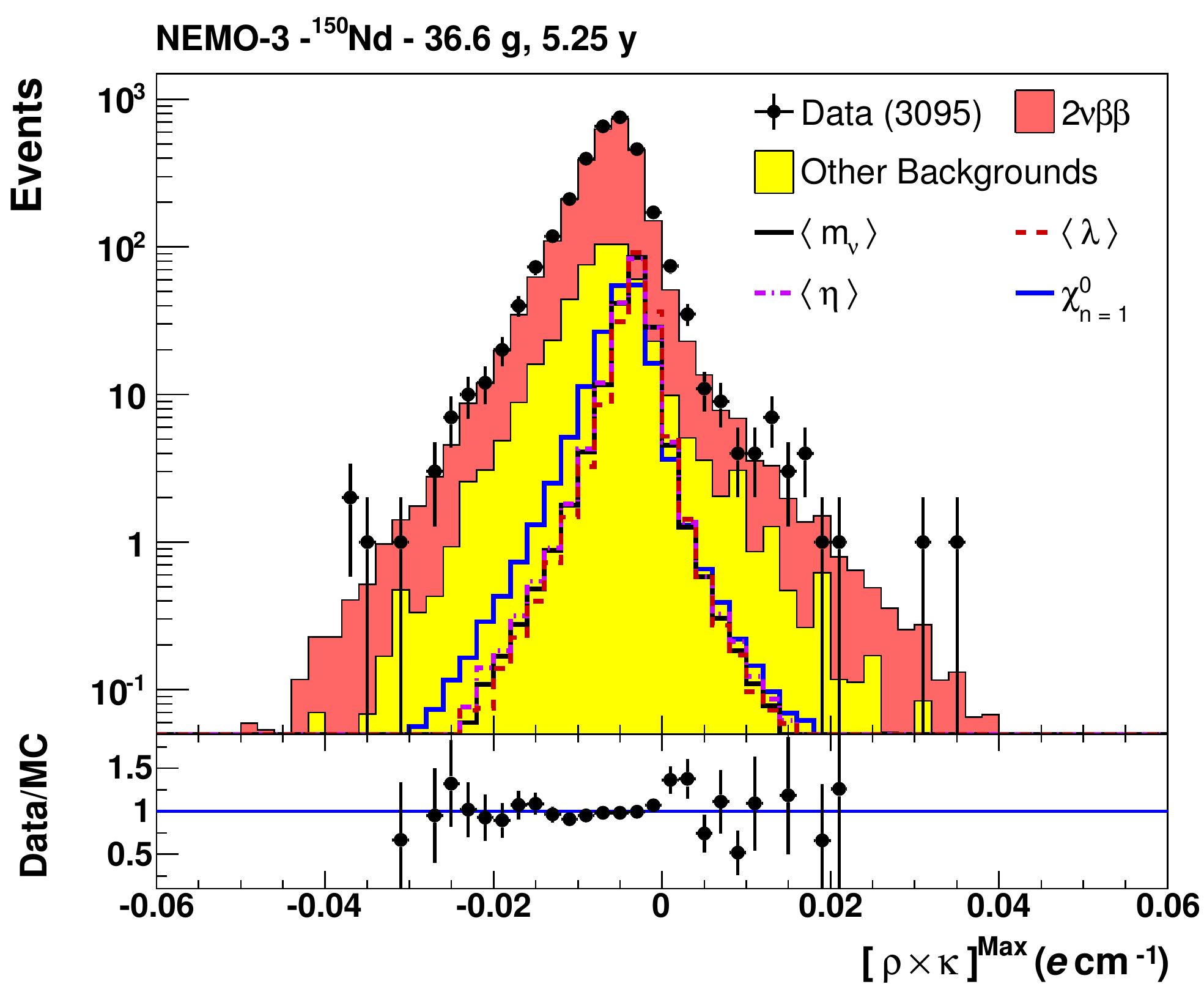}\put(40,160){\large\textbf{(e)}}\end{overpic}~\label{Fig:2e_CurvatureH}}%
    \subfigure{\begin{overpic}[scale=0.4]{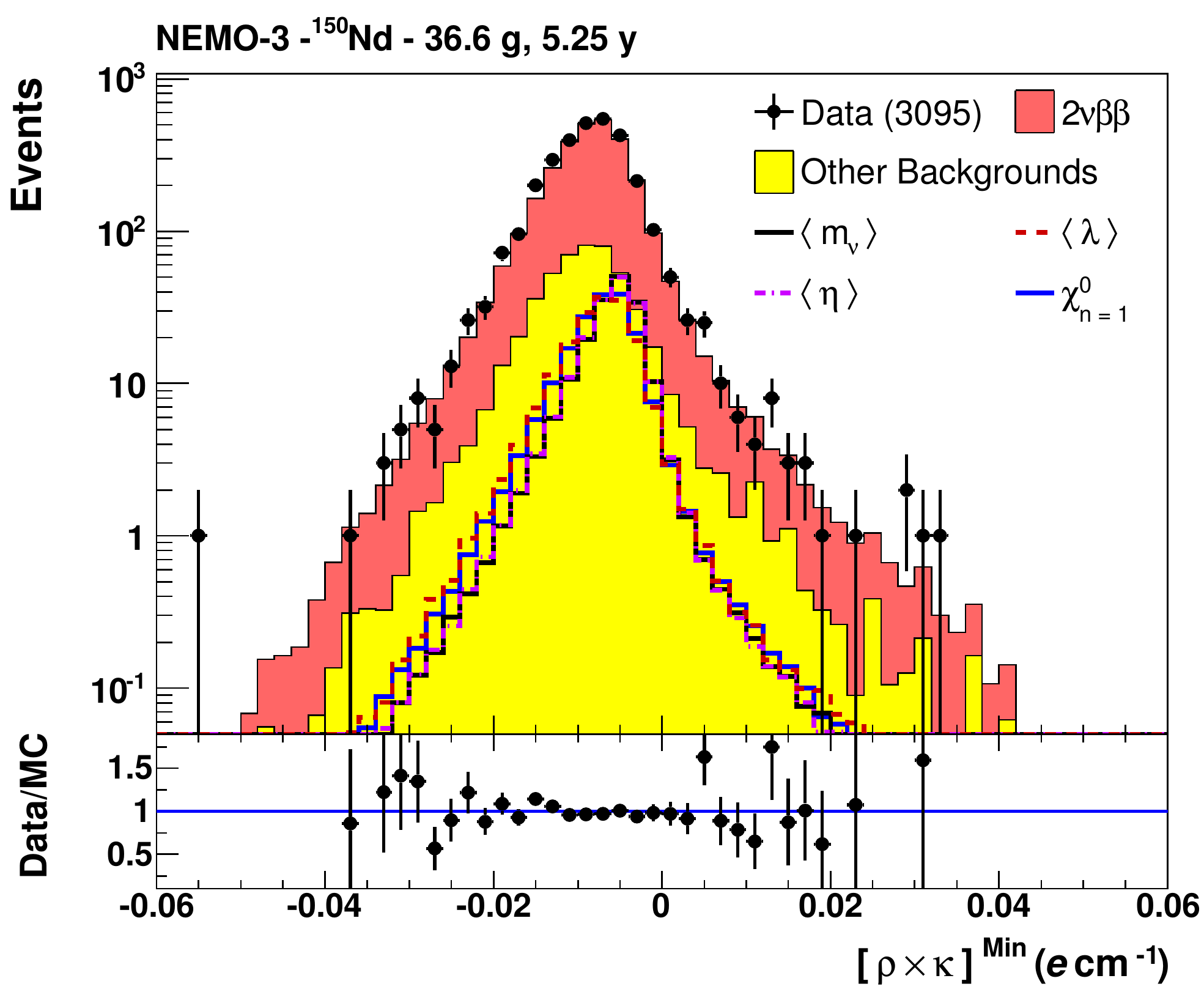}\put(40,160){\large\textbf{(f)}}\end{overpic}~\label{Fig:2e_CurvatureL}}%
   \caption{\label{Fig:2e_channel_observables2}  Distributions of~\subref{Fig:2e_Cosee} the opening angle between the two electron tracks, \subref{Fig:2e_Pint} the internal probability distribution, the track lengths associated with the \subref{Fig:2e_LeMax} higher and   \subref{Fig:2e_LeMin} lower energy electrons, and the signed track curvature ($\kappa\times\rho$) of the \subref{Fig:2e_CurvatureH} higher and \subref{Fig:2e_CurvatureL} lower energy electrons from the two-electron channel. The data are compared to the total expected background, and the \zeronu decay signals assuming several underlying mechanisms ($\langle m_{\nu} \rangle$, $\langle \lambda \rangle$, $\langle \eta \rangle$ and $\chi^{0}_{n = 1}$) are also shown with an arbitrary normalization in each figure, analogous to Fig.~\ref{Fig:2e_channel_observables1}.}
\end{figure*}

The total energy distribution of a single isotope does not discriminate between decays mediated by the mass mechanism and RHCs. However, the presence of a RHC in the electroweak Lagrangian results in a different angular distribution between the decay electrons when compared to the mass mechanism, and modifies their individual energy distributions as well~\cite{Ali:2007ec}. A unique feature of the \nemo detector among the current generation of $\beta\beta$-decay experiments is the ability to reconstruct the full topology of final state particles in each event, which provides an opportunity to determine the underlying decay mechanism, should \zeronu decay be observed~\cite{SuperNEMO}. This is demonstrated in Figs.~\ref{Fig:2e_channel_observables1} and~\ref{Fig:2e_channel_observables2}, where the differences between underlying \zeronu decay mechanisms are shown for several kinematic observables in the two-electron channel. 

Correlations between the observables can also be used to enhance the separation between the \zeronu decay signals and background, and thereby improve the sensitivity to \zeronu decays. This is demonstrated by employing a multivariate analysis in the two-electron channel in the search for \zeronu decay modes of $^{150}$Nd. A BDT analysis is performed using the TMVA package~\cite{TMVA} in ROOT~\cite{ROOT}.  The BDT is trained on ten observables that demonstrate good agreement between data and the MC simulation: the total energy of the two electrons ($E_{\mathrm{tot}}$); the higher ($E^{\mathrm{max}}_{e}$) and lower ($E^{\mathrm{min}}_{e}$) energy electron energies; the asymmetry between the electron energies defined as $A_{E} = (E^{\mathrm{max}}_{e} - E^{\mathrm{min}}_{e}) / E_{\mathrm{tot}}$; the track lengths associated with the higher ($L^{\mathrm{max}}_{e}$) and lower ($L^{\mathrm{min}}_{e}$) energy electrons; the opening angle between the two tracks ($\cos\theta$); the internal probability distribution ($P_{\mathrm{int}}$); and the curvature of the higher and lower energy electron tracks defined as $(\rho\times\kappa)^{\mathrm{max,min}}$, where $\rho$ is the charge in units of elementary electric charge $e$ and $\kappa$ is the reciprocal of the radius of track curvature. 
All observables are shown in Figs.~\ref{Fig:2e_channel_observables1} and~\ref{Fig:2e_channel_observables2} to demonstrate the separation between signal and background in each distribution.  

\begin{figure*}
    \subfigure{\begin{overpic}[scale=0.44]{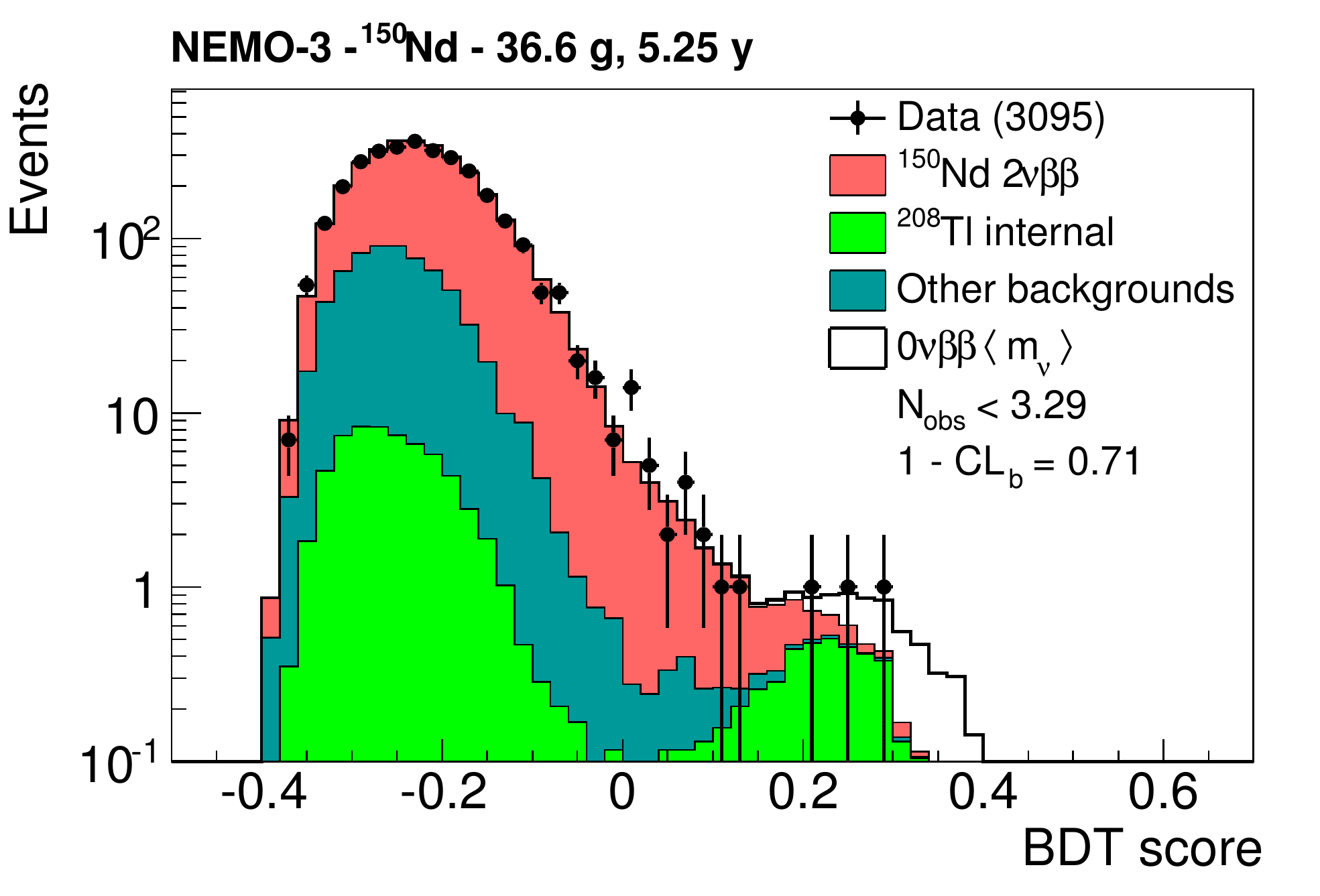}\put(40,140){\large\textbf{(a)}}\put(100,140){\textbf{Mass}}\put(100,130){\textbf{mechanism}}\end{overpic}\label{Fig:MM_limit}}%
    \subfigure{\begin{overpic}[scale=0.44]{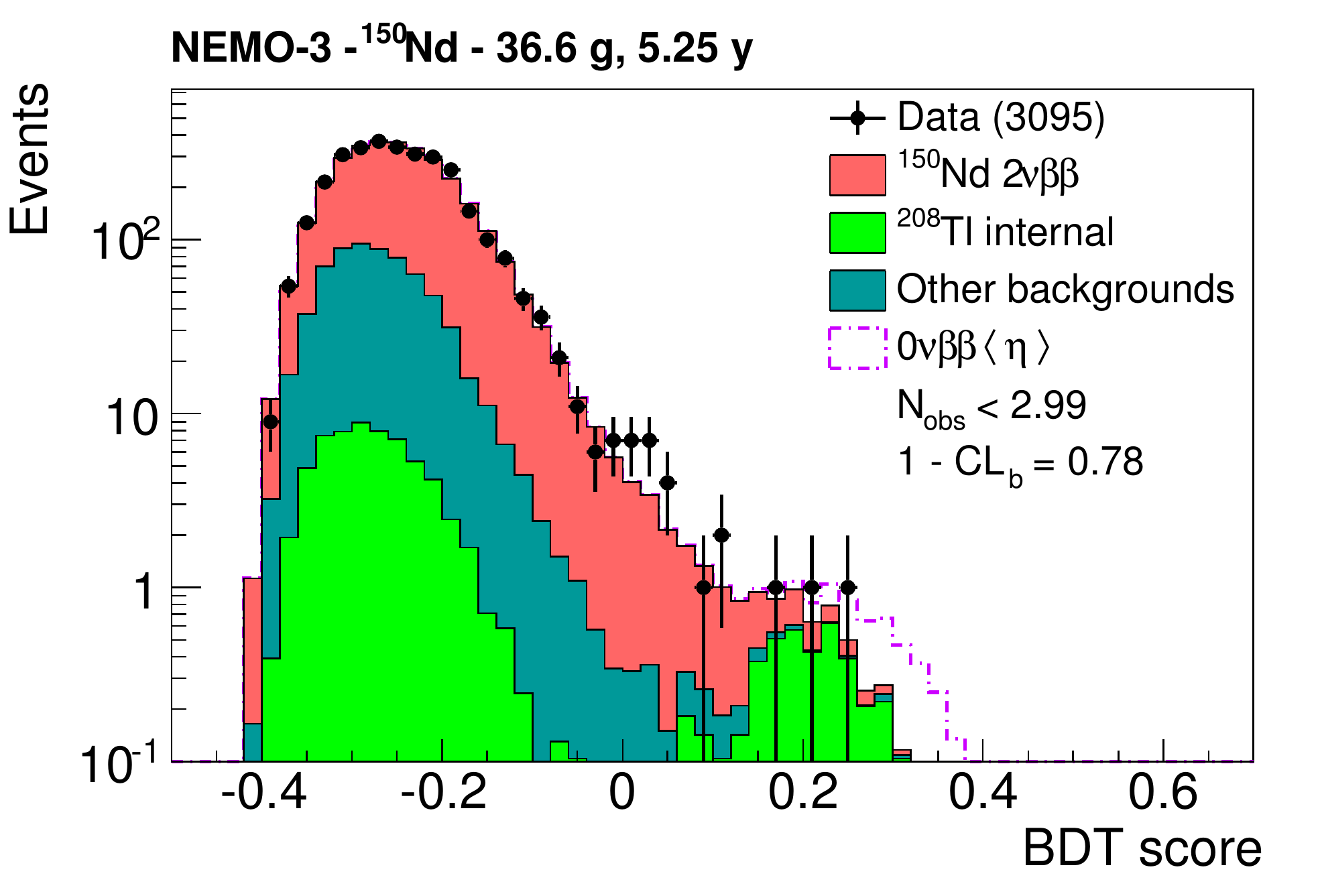}\put(40,140){\large\textbf{(b)}}\put(100,140){\textbf{RHC}}\put(100,130){\textbf{$\eta$ - mode}}\end{overpic}\label{Fig:ETA_limit}}\\
    \subfigure{\begin{overpic}[scale=0.44]{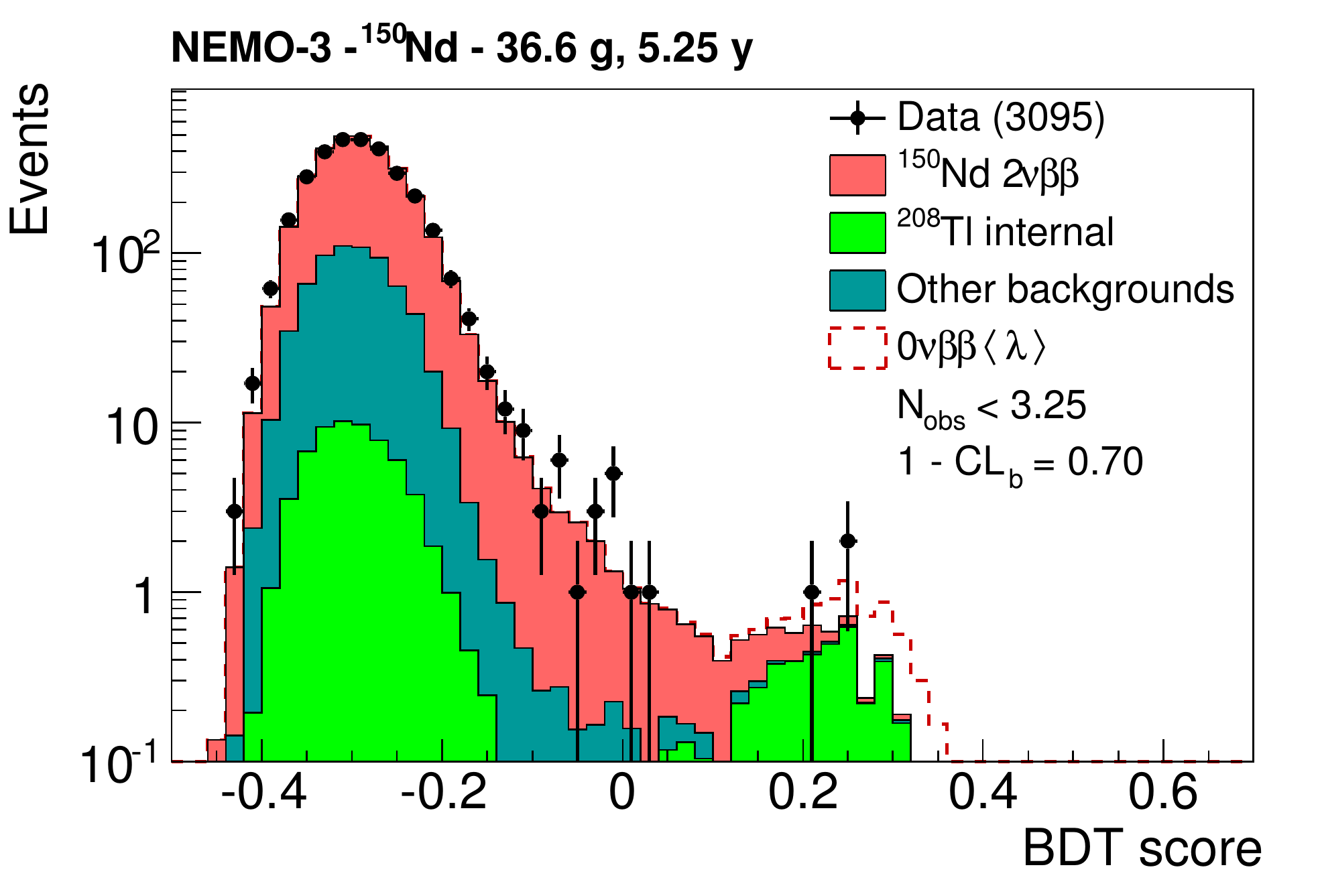}\put(40,140){\large\textbf{(c)}}\put(100,140){\textbf{RHC}}\put(100,130){\textbf{$\lambda$ - mode}}\end{overpic}\label{Fig:RHC_limit}}%
    \subfigure{\begin{overpic}[scale=0.44]{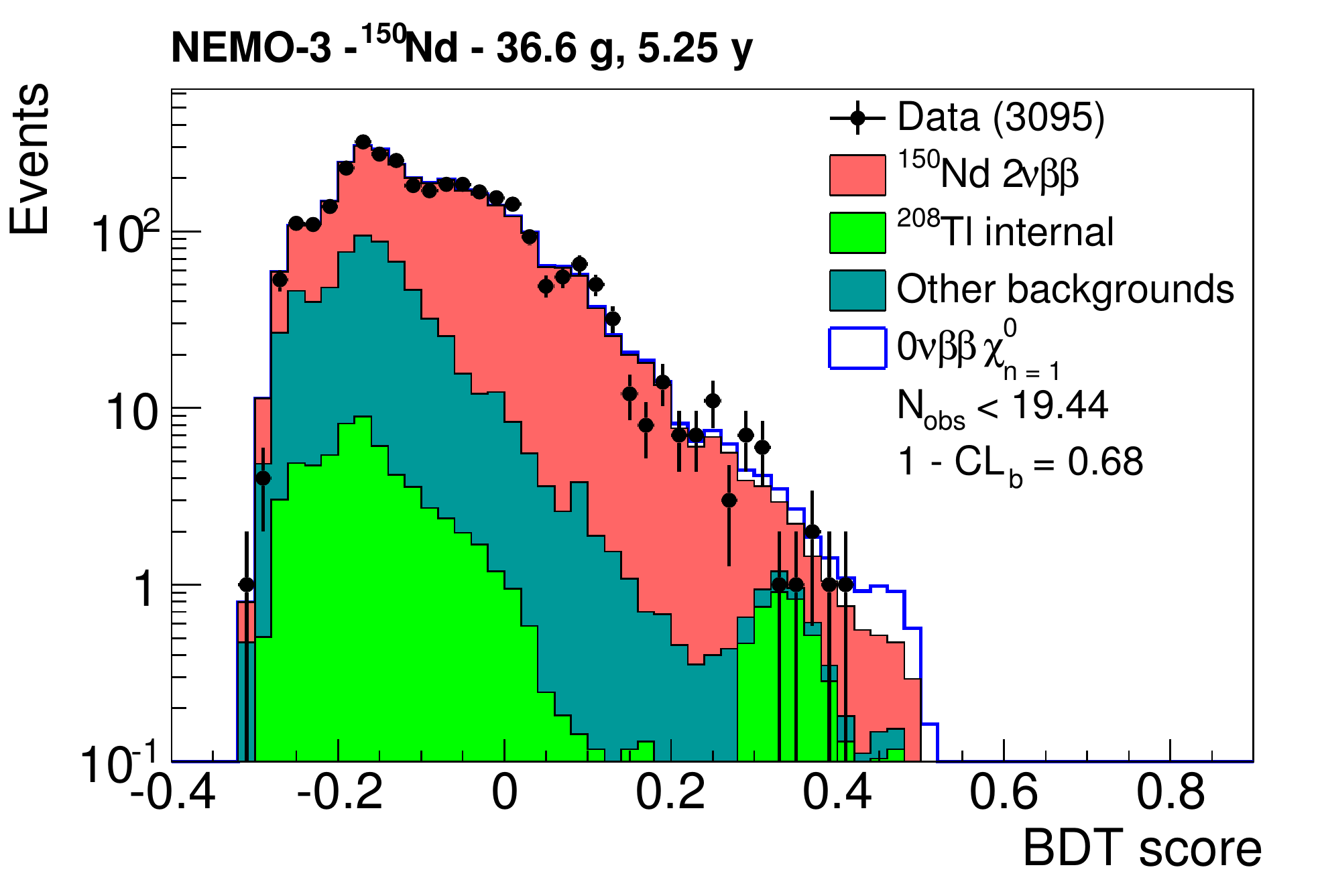}\put(40,140){\large\textbf{(d)}}\put(105,140){\textbf{Majoron}}\put(115,130){\textbf{n = 1}}\end{overpic}\label{Fig:MAJ1_limit}}
   \caption{\label{Fig:BDT_Limits} BDT score distributions resulting from the training for detecting \zeronu decays mediated via~\subref{Fig:MM_limit} a light Majorana neutrino exchange,~\subref{Fig:ETA_limit} RHCs coupling LH quarks to RH leptons,~\subref{Fig:RHC_limit} RHCs coupling RH quarks to RH leptons, and~\subref{Fig:MAJ1_limit} the emission of majorons with spectral index $n = 1$.  The data are consistent with the background-only hypothesis in each case as demonstrated by the observed $p$ values (1 - CL$_{b}$). The \zeronu decay rate normalized to the number of excluded events at the~$90\%$ C.L. for each mechanism is shown stacked on top of the expected background rate. 
}
\end{figure*}

Four BDTs, one for each of the \zeronu decay mechanisms considered, are trained using $20\%$ of the available MC statistics for each sample that contributes to the two-electron channel. The TMVA package provides various BDT parameters that can be tuned to yield a better sensitivity to the investigated process. The adaptive boosting algorithm with 850 trees yields good performance for each of the \zeronu decay mechanisms. The remaining $80\%$ of the MC statistics are used for testing the BDT performance. Each BDT algorithm returns a score that is a continuous variable distributed between $-1$, for more background-like events, and $+1$, indicating more signal-like events. The BDT scores of the testing samples for the mass mechanism, RHC mechanisms and majoron emission with spectral index $n = 1$ are shown in Fig.~\ref{Fig:BDT_Limits}. 

The most powerful discriminating variable in the BDT algorithms for all mechanisms investigated is $E_{\mathrm{tot}}$. The individual electron energies provide some additional sensitivity, as their correlations are different between many of the backgrounds and the signals. For example, the higher energy electron from $^{208}$Tl decays is primarily produced as a conversion electron with an energy of approximately \unit[2.5]{MeV}, while the other electron is a $\beta$ particle with an average energy of approximately \unit[0.56]{MeV}. In contrast, the electrons emitted in a \zeronu decay mediated by the mass mechanism share energy equally on average. By identifying these correlations, the BDT is able to improve the separation between $^{208}$Tl and $\langle m_{\nu} \rangle$ decays. The other observables act as weakly classifying variables in the BDT algorithm, with relatively similar discriminating power provided by each of them. The internal probability distribution is useful primarily in discriminating between signal events and those from decays of external backgrounds and radon in the tracker volume. The track length distributions also provide some discriminating power since they are correlated with the electron energies and their opening angle.

The BDT scores of the testing samples are used to search for evidence of \zeronu decay from $^{150}$Nd. A modified frequentist approach is employed, which uses a binned log-likelihood ratio test statistic (CL$_{s}$)~\cite{CLs,Read:2002hq,COLLIE}. The quantity CL$_{s}$ is defined as the ratio CL$_{s+b}$/CL$_{b}$, where CL$_{s+b}$ is the $p$ value of the data assuming the signal-plus-background hypothesis, and CL$_{b}$ is the $p$ value assuming the background-only hypothesis. To obtain a limit at the 90\% C.L., the signal strength is scaled until CL$_{s} = 1-0.9$. This method is useful in low background scenarios because taking the ratio mitigates the effects of under-fluctuating backgrounds that can otherwise lead to an overestimation of exclusion limits. Systematic uncertainties on background normalizations and signal efficiency are incorporated in the limit setting by fluctuating signal and background distributions by random numbers drawn from Gaussian functions with widths reflecting the magnitude of the systematic uncertainties. 

The systematic uncertainties considered for each of the background normalizations are the same as those previously described in Sections~\ref{Sec:Background_model}~and~\ref{Sec:2vBB_measurement}. The uncertainty on the normalization of the $^{150}$Nd \twonu decay distribution in the signal  region of interest is constrained by the statistical uncertainty of the measurement given in  Eq.~\ref{Eq:2vBB_halflife}, which is $2.4\%$. These uncertainties have a negligible effect on the expected sensitivity. The uncertainty on the \zeronu decay signal efficiency is assumed to be equal to the uncertainty on the \twonu decay efficiency. The dominant contribution  (5.55\%) again comes from the comparison of \nemo data from $^{207}$Bi calibration runs with HPGe measurements of the same calibration sources. This is added in quadrature with the uncertainties on the foil composition, enrichment, energy loss and bremsstrahlung simulation, which together yield an uncertainty of $[+5.8,-5.7]\%$ (see Table~\ref{Tab:2vBB_systematics}). Energy scale uncertainties that could affect the shape of the \zeronu signals have been shown to have an effect of $<1\%$ on the derived limits~\cite{Nd150_pub}.  

The data are consistent with the background-only hypothesis as indicated by the $p$ values ($1-\mathrm{CL}_{b}$) shown in Fig.~\ref{Fig:BDT_Limits} for the various \zeronu decay mechanisms investigated. Therefore, upper limits on the signal strengths are derived at the 90\% C.L. and translated into lower limits on the half-lives for each process. 

The total efficiency for selecting \zeronu events mediated by light Majorana neutrino exchange is $12.1\%$. The observed lower limit on the half-life for this process is \unit[$\mathrm{T}^{0\nu}_{1/2} > 2.0 \times 10^{22}$]{y}, which is consistent with the $\pm 1$ standard-deviation range of the median expected limit as shown in Table~\ref{Tab:0vBB_limits}. This is the most stringent limit obtained for the isotope $^{150}$Nd assuming the mass mechanism. We convert it into an upper limit on the effective neutrino mass $\langle m_{\nu} \rangle$ using the phase space value of \unit[$G^{0\nu} = 63.03 \times 10^{-15}$]{y$^{-1}$}~\cite{Kotila:2012zza}, nuclear matrix elements spanning the range $M^{0\nu}=1.71 -5.60$~\cite{Chaturvedi:2013,Rodriguez:2010,Fang:2011,Terasaki:2015,Barea:2015kwa,Song:2014,Mustonen:2013}, and the axial vector coupling $g_{A} = 1.27$~\cite{Agashe:2014kda}. The corresponding mass limit is \unit[$\langle m_{\nu}\rangle < \lbrack 1.6 - 5.3 \rbrack$]{eV}. The range is due to the large differences in $M^{0\nu}$ values calculated with different methods. 
We only use $M^{0\nu}$ calculations that take into account the difference between the deformation of the initial ($^{150}$Nd) and final state ($^{150}$Sm) nuclei.

The BDT technique improves the expected (observed) half-life limit for the mass mechanism process by $11\% (34\%)$ compared to the results obtained using only the total energy distribution. This is primarily due to the improved ability to discriminating between $^{208}$Tl decays and the $\langle m_{\nu} \rangle$ signal. Since the sensitivity for $T^{0\nu}_{1/2}$ scales with the square root of the exposure, this improvement corresponds to an increase by a factor of $\approx 1.2$ in exposure.

The half-life limit for the light Majorana neutrino exchange can also be used to derive limits on the 
coupling parameter $\lambda^{'}_{111}$ of $R$-parity violating ($\cancel{R}_{p}$) supersymmetry (SUSY) models. Under the assumption that the decay proceeds via a short-range gluino or neutralino exchange, the kinematics are sufficiently similar to the mass mechanism such that the same signal efficiency and signal template can be used to search for $\cancel{R}_{p}$ decays. Using the same half-life limit of \unit[$\mathrm{T}^{0\nu}_{1/2} > 2.0 \times 10^{22}$]{y}, a phase space factor of \unit[$G^{0\nu} = 63.03$]{$\times 10^{-15}$y$^{-1}$} and nuclear matrix elements from Ref.~\cite{Faessler:1998qv}, we obtain an upper limit on the coupling constant of $\lambda^{'}_{111} \leq 1 \cdot 10^{-4} \times f$, where 
\begin{equation}
\begin{small}
f = C\times\left(\frac{m_{1}}{1~\mathrm{TeV}}\right)^{2}\left(\frac{m_{2}}{1~\mathrm{TeV}}\right)^{1/2}.
\end{small}
\end{equation}
In the case of gluino exchange $C = 1.8$, $m_{1}$ is the squark and $m_{2}$ the gluino mass. If the process is mediated by neutralinos, $C = 12.5$, $m_{1}$ is the selectron and $m_{2}$ the neutralino mass~\cite{Faessler:1998qv}.

The observed lower limits on the half-lives for decays mediated by RHCs are $T^{0\nu}_{1/2}>$~\unit[1.9$\times10^{22}$]{y} for the $\langle \eta \rangle$ mode and $T^{0\nu}_{1/2}>$~\unit[1.1$\times10^{22}$]{y} for the $\langle \lambda \rangle$ decay mode.  These are the most stringent limits to date obtained with the isotope $^{150}$Nd. An improvement of $14\% (48\%)$ in the expected (observed) limit is obtained through the use of a BDT for the decays mediated by the $\langle \eta \rangle$ decay mode. A $7\% (35\%)$ improvement is achieved for the expected (observed) limits on the half-life assuming the $\langle \lambda \rangle$ mode. The improvement in the expected sensitivity is smaller than for the $\langle m_{\nu} \rangle$ and $\langle \eta \rangle$ decay modes since the kinematic distributions of $^{208}$Tl decays and the $\langle \lambda \rangle$ mode are similar. 

We also derive a limit of $T^{0\nu}_{1/2}>$~\unit[0.3$\times10^{22}$]{y} (observed)
for the emission of a majoron with spectral index $n = 1$. 
While the use of the BDT improves the sensitivity by approximately $5\%$, the observed limit shows only a $2\%$ improvement over the univariate method. The half-life can be converted into a limit on the neutrino-majoron coupling strength of $\langle g_{ee} \rangle < (3.8 - 14.4)\times 10^{-5}$, using $G^{0\nu\chi^{0}_{n=1}} = 3.1\times 10^{-15}$~y$^{-1}$~\cite{Kotila:2015ata}, $g_{A} = 1.27$, and  
the same values of $M^{0\nu}$ as used for the mass mechanism. This is the best limit on $\langle g_{ee} \rangle$ obtained with $^{150}$Nd. 

\begin{table}
	\centering
	\begin{tabular}{c|c|c|ccc}
	\hline
	\hline
		\multirow{2}{*}{Mechanism} & \multirow{2}{*}{$\varepsilon$ (\%)} & $T^{0\nu}_{1/2}$ & \multicolumn{3}{c}{$T^{0\nu}_{1/2}$} (Expected) \\
		 & & (Observed) & -1$\sigma$ & median & +1$\sigma$\\
		\hline 
        $\langle m_{\nu}\rangle$, $\cancel{R}_{p}$ SUSY & 12.1 & 2.0 & 1.3 & 1.8 & 2.3 \\
         $\langle \eta \rangle$& 10.4 & 1.9 & 1.1 & 1.5 & 2.0 \\
         $\langle \lambda \rangle$& 6.8 & 1.1 & 0.7 & 1.0 & 1.2\\
        $\chi^{0}_{n=1}$ & 9.4 & 0.3 & 0.1 & 0.2 & 0.3 \\
         \hline
         \hline
	\end{tabular} 
    \caption{\label{Tab:0vBB_limits} Observed lower limits on $T^{0\nu}_{1/2}$ (in units of \unit[10$^{22}$]{y}) at the 90\% C.L.~for all \zeronu decay mechanisms investigated in this analysis. 
   The median expected limits and the $\pm 1$ standard deviation ($\sigma$) bands are also given. 
     }
\end{table}

\section{Conclusions}
\label{Sec:Conclusions}

We use the entire \nemo data set to obtain a complete background description of the $^{150}$Nd source foil. All internal and external background decay rates are measured using multiple dedicated analysis channels, which are defined by the final-state topology of each event to enhance the sensitivity to the isotopes of interest.  With a live time of 5.25 years and \unit[36.6]{g} of $^{150}$Nd, the \twonu decay half-life for the ground state transition is measured to be $T^{2\nu}_{1/2} =$~\unit[$\lbrack 9.34 \pm 0.22~\mathrm{(stat)}~^{+0.62}_{-0.60}~\mathrm{(syst)}\rbrack$]{$\times 10^{18}$y}.  This result represents the most accurate measurement of the half-life of $^{150}$Nd to date.

A BDT multivariate technique is used to maximize the sensitivity to various \zeronu signals from different BSM physics models. This is the first instance that a multivariate analysis has been employed in this way to derive limits on \zeronu decay processes. As there is no indication of \zeronu decay, lower limits are derived for each mechanism.  The lower limit on the half-life obtained assuming the mass mechanism is \unit[$T^{0\nu}_{1/2} > 2.0 \times 10^{22}$]{y} at the $90\%$~C.L.  This corresponds to an upper limit on an effective neutrino mass of $\langle m_{\nu} \rangle <$~\unit[$\lbrack 1.6 - 5.3 \rbrack$]{eV}.  The use of a BDT technique to increase the discrimination between signal and background improves the expected sensitivity by approximately $11\%$ compared to the results obtained using only the total energy distribution of the two electrons. This demonstrates the power of both the combined tracking and calorimetric detector design adopted by the \nemo experiment and the multivariate analysis technique. 

This article presents the most stringent half-life limits on several \zeronu decay mechanisms using the isotope $^{150}$Nd. Due to the small mass of this sample, the limits are in general not competitive with the leading results in this field~\cite{Arnold:2015wpy, Tl130_1, Agostini:2013mzu,Gando:2012zm}. The exception is the limit for the coupling between neutrinos and majorons $\langle g_{ee} \rangle$ with $n = 1$.  The limit on $\langle g_{ee} \rangle$ is comparable with results obtained using other isotopes with significantly larger exposures~\cite{Arnold:2015wpy, Tl130_1, Ge76_2,PhysRevC.86.021601}. This is due to the favourable nuclear properties of $^{150}$Nd, including its large $Q_{\beta\beta}$ value and atomic number, as well as the excellent background reduction possible through the selection of two-electron events with the \nemo detector.

The results presented herein are promising for the SuperNEMO experiment, which is based on the same design principles as the \nemo detector, and will have much lower backgrounds and improved energy resolution.  The demonstrator module for SuperNEMO is currently under construction at the LSM laboratory. The main isotope used to search for \zeronu decay will be $^{82}$Se with a total mass of 7 kg. However, there is potential to also include $^{150}$Nd and $^{48}$Ca as secondary isotopes.  
The analysis presented in this article demonstrates that $^{150}$Nd is indeed an attractive isotope for future \zeronu decay searches, assuming an increase in the exposure.  In addition, it has been shown that the use of a multivariate analysis technique, such as a BDT, has the potential to greatly improve the sensitivity for experiments that can measure multiple observables in the final state.

\section*{Acknowledgements}
The authors would like to thank the staffs of the Modane Underground Laboratory for their technical assistance in operating the detector. We acknowledge support by the funding agencies of the Czech Republic, the National  Center  for Scientific Research/National  Institute  of  Nuclear  and  Particle
Physics (France), the Russian Foundation for Basic Research (Russia), the Science and Technology Facilities Council (United Kingdom), and the National Science Foundation (United States).

\end{document}